\documentclass[20pt]{article}

\usepackage{amsmath}
\usepackage{graphicx}
\usepackage{amsmath,amssymb,mathtools}
\usepackage{graphics,color,array,calc,rotating,epsfig,psfrag}
\numberwithin{equation}{section}
\usepackage[super,compress]{cite}
\usepackage{bm}
\usepackage{dcolumn}
\usepackage{float}
\usepackage{subfigure}
\usepackage{url}
\usepackage{subfloat}
\usepackage[super,compress]{cite}
\usepackage{hyperref}
\hypersetup{colorlinks,urlcolor=black,citecolor=black,linkcolor=black,filecolor=black}
\usepackage{breakurl}
\begin{document}

\begin{center}
{\textbf{\large Neutrino spin oscillations in gravitational fields in noncommutative higher dimensions }}\\

{\small S.A. Alavi$^{\dag}$ $^{\ddag}$,T. Fallahi.Serish$^{\star}$$^{\ast}$\\

$^{\dag}$\small \emph{Department of physics, Hakim Sabzevari University, P.O. Box 397}\\
\small \emph{Sabzevar, Iran}\\
$^{\star}$\small \emph{Faculty of Physics, Shahrood University of Technology,
P.O.Box 3619995161 Shahrood, Iran}\\
$^{\ddag}$\small \emph{s.alavi@hsu.ac.ir; alaviag.gmail.com}}\\
$^{\ast}$\small \emph{fallahi.ta@shahroodut.ac.ir}
\end{center}

\begin{abstract}
Investigation of neutrino spin oscillation in the gravitational fields of Black Holes (BH) is one of the interesting  topics in neutrino physics. On the other hand, in recent years, many studies have been devoted to the  exploration of different physical phenomena in higher dimensions. Noncommutative geometry has also  been focus of researchers  in recent years to explore the structure of space-time more deeply and accurately. In this work, the neutrino spin oscillation in the noncommutative higher-dimensional  gravitational fields of  Schwarzschild  and Reissner-Nordström metrics  is studied. The effects of non-commutativity of space are calculated and its role in different dimensions is discussed. Considering the fact that in the non-commutative space, for certain values of $\frac{M}{\sqrt{\theta}}$, no event horizon exists and, consequently, there will be no black hole, some upper bounds for the non-commutative parameter is obtained.
\end{abstract}
\large
\section{Introduction}

Since the first time in 1930 that Pauli postulated the existence of neutrinos to establish the conservation of energy, until now, the discussion and research about this particle and its properties \cite{1} have made us understand matter better, but the field of neutrino physics is still surrounded by more questions than answers. What is the mechanism behind neutrino masses? Which neutrino is the lightest? How many flavors of neutrinos are there? Do neutrinos interact with matter or among themselves in a way that differs from the Standard Model prediction? Are neutrinos connected to the dark matter or the dark energy content of the universe? Are neutrinos their own antiparticles? Do neutrinos violate the symmetries of physics? Besides those theoretical questions, some experimental challenges exist too.  At present, the study of neutrinos from different sources and at different energies using cutting-edge technologies is entering new eras of innovation. Now, neutrino is an important basis of various astrophysical and cosmological researches.
On the other hand, since Einstein’s proposal of general relativity (GR) in 1915, a lot of research has been devoted to unify (GR) and Electromagnetism as two fundamental interactions in nature. In addition, there were some efforts in 1920s to unify these interactions through Kaluza-Klein theory, which was a classical unified field theory built in five-dimensional space-time\cite{2,3}. Recently, motivated by string theory as a requirement for describing a consistent theory of quantum gravity, extra dimensions have been the subject of much attention. Besides string theory, there are some other theories proposing the necessity of extra dimensions. Large extra dimensions, mostly motivated by the ADD model, by Arkani-Hamed, Dimopoulos, and Dvali together with Antoniadis in Refs. \citen{4,5} to solve the hierarchy problem. While the electromagnetic, weak and strong interactions differ by just six orders of magnitude, the gravitational interaction falls apart by  a further thirty-three orders\cite{5}. Warped extra dimensions were proposed by the Randall-Sundrum (RS) model \cite{2,6}, in which our observable universe is modeled as a four-dimensional hypersurface, known as the 3-brane, embedded in a five dimensional space, usually called the bulk. The novel idea of the brane world is that all the gauge interactions, described by the Standard Model are confined to live in the 3-brane while the gravitational interaction can spread into the fifth dimension of the space. Universal extra dimensions,were proposed and first studied in Ref. \citen{7}.Motivated by different sources, the formulation of physical theories in more than four space-time dimensions is a subject that has received increasing attention in recent years. On the other hand, it is broadly accepted that the idea of continuous space-time should break down at very short distances.  It has been shown that non-commutative geometry naturally appears in string theory with a non-zero antisymmetric B field\cite{8,9}. They studied the behavior of open strings in the presence of a constant B-field and showed that the magnetic field(B) dependence of the effective action is completely described by making space-time non-commutative. It is also shown that a string theory membrane wrapped around a torus in the presence of background B field, manifests non-commutative coordinates as a simple consequence of canonical commutation relations. They further showed that by taking B large, holding the effective open string parameters fixed, one can get an effective description of the physics in terms of the non-commutative Yang-Mills theory. This means that there must be a change of variables from ordinary to non-commutative Yang-Mills fields.  Therefore reformulations of  theories on non-commutative spaces may play an important role in explaining the properties of nature at the Planck scale. \\

Furthermore, the interaction of neutrinos with gravitational fields is one of the interesting phenomena that can lead to the transition between different states of helicity (spin oscillation).  It can also lead to neutrino flavor oscillations, i.e., neutrinos of one type can be converted into another type if there is a mixing between different neutrino eigenstates. Examples of this oscillation type are $\nu_{e}\rightarrow\nu_{\mu}$ or $\nu_{\mu}\rightarrow\nu_{\tau}$ conversions that are the possible explanations of the solar and atmospheric neutrino problems. The electromagnetic interactions are of great importance in studying neutrino spin and spin-flavor oscillations. For example, the neutrino interaction with an external electromagnetic field provides one of the mechanisms for the mixing between different helicity eigenstates. It is also interesting to study the influence of gravitational fields on neutrino oscillations. Despite the fact that,  the gravitational interaction is relatively weak compared to electromagnetic interaction, there are rather strong gravitational fields in the Universe i.e., gravitational fields of black holes. Neutrino oscillations in gravitational fields are of interest in astro-particle physics and cosmology. The influence of the gravitational interaction on neutrino oscillations has been studied in many publications (see, e.g., Refs. \citen{10,11,12,13,14,15,16}).
In this paper we study neutrino spin oscillations in non-commutative higher-dimensional gravitational fields. We investigate how the non-commutativity of space affects the neutrino spin oscillations in three and higher dimensions. We also provide upper bounds on noncommutative parameter $\theta$.

\section{Neutrino spin oscillations in non-commutative gravitational fields}
The non-commutative  space coordinates $x^{\mu}$  satisfy the following commutation relations:\cite{17}

\begin{equation}\label{1l}
[x^{\mu},x^{\nu}]=i\theta^{\mu\nu},[x^{\mu},p^{\nu}]=i\delta^{\mu\nu},[p^{\mu},p^{\nu}]=0.
\end{equation}
The simplest case is that $\theta^{\mu\nu}$  be a constant  antisymmetric tensor with real elements which have  square length dimension. To study gravitational fields in non-commutative spaces, it is not necessary to change the part of the Einstein tensor in the field equations, and non-commutative effects only act on the matter source. Non-commutativity of space implies to replacing the point-like structures by smeared objects. So we  change the point distribution like source to spread objects. The spread effect is implemented mathematically as follows: Dirac delta function everywhere  is replaced by a  Gaussian distribution of  minimum width $\sqrt{\theta}$. Therefore, we choose the mass  and charge densities  of a static, spherically symmetric, smeared, particle-like gravitational source as (see, Refs. \citen{17,18,19}):
\begin{equation}\label{2l}
  \rho_{\theta}(r)=\frac{M}{(4\pi\theta)^{\frac{3}{2}}}exp(-\frac{r^{2}}{4\theta}),
\end{equation}

\begin{equation}\label{3l}
  \rho_{Q}(r)=\frac{Q}{(4\pi\theta)^{\frac{3}{2}}}exp(-\frac{r^{2}}{4\theta}).
\end{equation}

A particle with mass M instead of being completely in one spot, spread out through a region of space with the area  $\theta$.  For an observer at large distance this smeared density looks like a small sphere of matter with radius about $\sqrt{\theta}$ of mass $M$ and charge $Q$, so the metric becomes Schwarzschild ($Q=0$) or RN black hole at large distances. solutions of Einstein equations with such smeared sources give new kind of regular black holes in four (Refs. \citen{19,20,21,22,23,24}) and higher dimensions (Refs. \citen{25,26,27,28,29,30,31,32,33}). \\
Vierbein four-velocity $u^{a}=(u^{a},u_{1},u_{2},u_{3})$ and four-velocity of the particle in its geodesic path $U^{\mu}$ are related by the following relation\cite{34,35,36,37}:

\begin{equation}\label{4l}
u^{a}=e^{a}_{\mu}U^{\mu},
\end{equation}
where $e^{a}_{\mu}$ are non-zero vierbein vectors and we can decompose the four-velocity vectors $U^{\mu}$  using spherical coordinates ${U^{\mu}}=(U^{0},U_{r},U_{\theta},U_{\varphi})$  which  its components are defined as follows:
\begin{equation}\label{5l}
U^{0}=\frac{dt}{d\tau},U_{r}=\frac{dr}{d\tau},U_{\theta}=\frac{d\theta}{d\tau},U_{\varphi}=\frac{d\varphi}{d\tau}.
\end{equation}

 $ U^{\mu} $  \textrm{is related to the world velocity of the particle through} $ \overrightarrow{U}=\alpha\overrightarrow{v} $  \textrm{where} $ \alpha=\frac{dt}{d\tau} $ \textrm {and} $\tau$ \textrm{is the proper time. To study the spin oscillation of a particle in a gravitational field we need to calculate } $ G_{ab}=(\overrightarrow{E},\overrightarrow{B}) $ \textrm{which is the analog tensor of the electromagnetic field and  is defined as\cite{37}:}

\begin{equation}\label{6l}
G_{ab}=e_{a\mu;\nu}e^{\mu}_{b}U^{\nu},G_{0i}=E_{i},G_{ij}=-\varepsilon_{ijk}B_{k}.
\end{equation}

\textrm {$ e_{a\mu;\nu} $ are the covariant derivatives of vierbein vectors:}

\begin{equation}\label{7l}
e_{a\mu;\nu}=\frac{\partial e_{a\mu}}{\partial_{x^{\nu}}}-\Gamma^{\lambda}_{\mu\nu}e_{a\lambda},
\end{equation}

\textrm{ where $ \Gamma^{\lambda}_{\mu\nu} $ is the Christopher symbol. Geodesic equation of a particle in a gravitational field is:\cite{38,39}}

\begin{equation}\label{8l}
\frac{d^{2}x^{\mu}}{dq^{2}}+\Gamma^{\mu}_{\sigma\upsilon}\frac{dx^{\sigma}}{dq}\frac{dx^{\upsilon}}{dq}=0,
\end{equation}

\textrm{where the variable $ q $ parameterizes the particle’s world line.}

\textrm{Neutrino spin oscillations frequency is given by the expression} $ \overrightarrow{\Omega}=\frac{\overrightarrow{G}}{\alpha} $ \textrm{, where vector} $ \overrightarrow{G}$ \textrm{is defined in this way:}\cite{37}

\begin{equation}\label{9l}
\overrightarrow{G}=\frac{1}{2}(\overrightarrow{B}+\frac{1}{1+u^{0}}[\overrightarrow{E}\times\overrightarrow{u}]).
\end{equation}

\section{ Neutrino spin oscillations in non-commutative higher dimensional Schwarzschild metric}

In  this section, by solving Einstein's equations with $\rho_{\theta}(r)$ as a source of matter and using  the
natural units $\hbar=c=1$, we study stationary, spherically symmetric, asymptotically flat metrics in $(d+1)$ dimensions that is described by the most general standard form of the space-time element:

\begin{equation}\label{10l}
ds^{2}=A^{2}(r)dt^{2}-A^{-2}(r)dr^{2}-r^{2}d\Omega^{2}_{d-1},
\end{equation}
where d is the dimension of space and $d\Omega^{2}_{d-1}$ denotes the differential solid angle in d  dimensions . For the Schwarzschild black hole with mass $M$ in three dimensional space ($d=3$) we have:

\begin{equation}\label{11l}
ds^{2}=A^{2}(r)dt^{2}-A^{-2}(r)dr^{2}-r^{2}(d\theta^{2}+\sin^{2}\theta d\phi^{2}),
\end{equation}
with :
\begin{equation}\label{12l}
A(r)=1-\frac{2MG}{r},
\end{equation}
 where $M=\frac{r_{g}}{2}$ and  $r_{g}$ is the gravitational radius.
the tt component of this metric in the non-commutative space is defined as follows \cite{18}:
\begin{equation}\label{13l}
  A(r)=\sqrt{1-\frac{4 G M}{\pi^{\frac{1}{2}}r}\gamma(\frac{3}{2},\frac{r^{2}}{4\theta})}.
\end{equation}

$ G $ is the Newton's constant with dimension $L^{d-1}$ and we define the dimensionless parameter $\eta$  as:

\begin{equation}\label{14l}
  \eta=\frac{M}{\sqrt{\theta}}.
\end{equation}

\begin{figure}[H]
\centering
\includegraphics[width=3.5in]{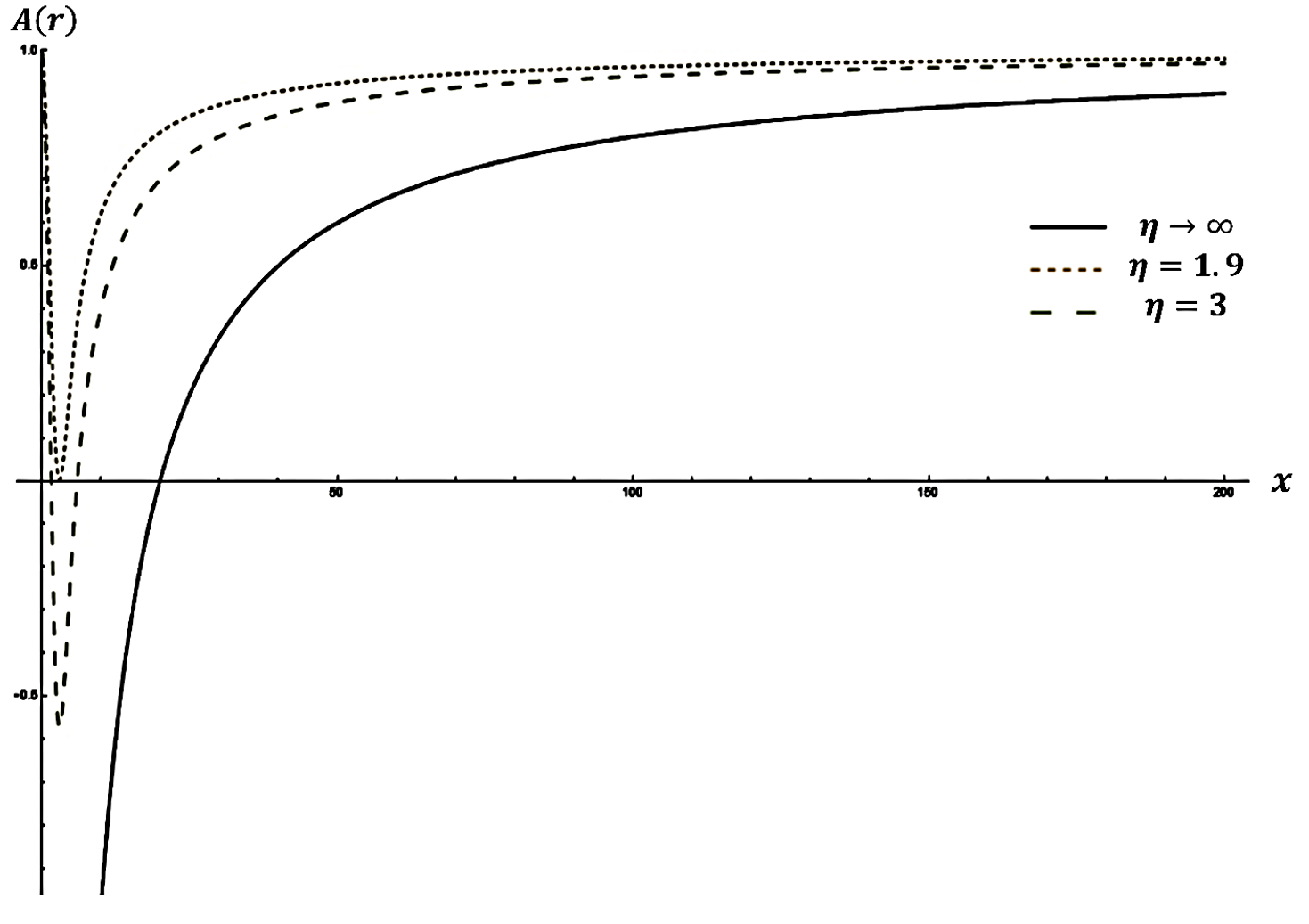}
\caption{\scriptsize The $A(r)$ component of the Schwarzschild metric vs. $ x $ for different values of  $\eta$ in 3 dimensions.\label{Fig0}}
\end{figure}

 By setting $Q=0$ in the time component of the RN metric in NC higher dimensions in the Ref. \citen{32}, the time component of the Schwarzschild metric is obtained, so for higher dimensions, the above metric is given as follows:
\begin{equation}\label{15l}
  A(r)=\sqrt{1-\frac{4 G M}{\pi^{\frac{d-2}{2}}r^{d-2}}\gamma(\frac{d}{2},\frac{r^{2}}{4\theta})},
\end{equation}.

where $ \gamma(\frac{d}{2},\frac{r^{2}}{4\theta})= \int^{\frac{r^{2}}{4\theta}}_{0} t^{\frac{d}{2}-1}e^{-t}dt $. \\
We define dimensionless variable $x=\frac{r}{\sqrt{\theta}}$ and plot $A(r)$ in three dimensions as a function of $x$ for different values of $\eta$ in Fig. \ref{Fig0}.

The location of the event horizon is calculated by the equation $A(r)=0$.  The existence of two intersection points with the horizontal axis indicates the formation of a black hole with two event horizons, while the presence of a single intersection suggests a degenerate horizon.

According to Eq.(\ref{14l}) $\eta$ is inversely proportional to the noncommutative parameter $\theta$.  In NC space for $\eta=\frac{M}{\sqrt{\theta}}>1.9 $ there are two distinct horizons shown by dash in Fig. \ref{Fig0} and for $\eta=\frac{M}{\sqrt{\theta}}=1.9 $ there is a degenerate horizon shown by dot. The case $\eta\longrightarrow\infty$  shows the metric in commutative space. As observed from the figure, by increasing the noncommutative parameter $\theta$ (decreasing $\eta$), the metric becomes flat in shorter distances  which indicates that, the non-commutativity of space has an opposite role to the curvature of space.\\

Vierbein components of vector $u^{a}$ is defined as follows:

\begin{equation}\label{16l}
   u^{a}=(\alpha A , U_{r}A^{-1}, U_{\theta}r, U_{\phi}r Sin\theta),
 \end{equation}

 since the Schwarzschild metric is spherically symmetric, we can assume that the neutrino moves in the equatorial plane $\theta=\frac{\pi}{2}$, i.e. $d\theta=0$. The magnetic field is given by $ \vec{B}=(U_{\phi}Cos\theta, -U_{\phi}A Sin\theta, U_{\theta}A) $ and for the orbits that lie in the equatorial plane we have  $\\{\upsilon}_{\theta}=0$ and $U_{\theta}=0$. $\Omega_{1}$  and   $\Omega_{3}$  depend on $\cos\theta$  and  $\\{\upsilon}_{\theta}$  so  $\Omega_{1}=\Omega_{3}=0$. The four-vector vierbein, the electric-gravity and magneto-gravity fields take the following  forms respectively\cite{34,37}:

 \begin{equation}\label{17l}
  u^{a}=(\alpha A,0,0,\alpha \\{\upsilon}_{\varphi}r),
\end{equation},

\begin{equation}\label{18l}
  E=(E_{0},0,0),B=(0,-\\{\upsilon}_{\varphi}A\alpha,0),
\end{equation}

$E_{0}$, ${\upsilon}_{\varphi}$ and $\alpha^{-1}$ are defined as follows:\cite{38}
\begin{equation}\label{19l}
v_{\varphi}=\frac{d\varphi}{dt}=\sqrt{\frac{F(r)}{2r}},
\end{equation}

\begin{equation}\label{20l}
E_{0}=\frac{1}{A^{2}(r)}(1-\frac{rF(r)}{2A^{2}(r)}),
\end{equation}

\begin{equation}\label{21l}
\alpha^{-1}=\frac{d\tau}{dt}=\sqrt{A^{2}(r)-\frac{1}{2}rF(r)},
\end{equation}

\textrm{where} $F(r)=\frac{dA^{2}(r)}{dr}$.\\
We can obtain $\Omega_{2}$ which is the only nonzero component of the neutrino spin oscillations frequency  in Schwarzschild metric by inserting time metric component A(r) in Eqs. (\ref{17l}) to (\ref{21l}), then by substituting Eq.(\ref{9l}) in $\overrightarrow{\Omega}=\frac{\overrightarrow{G}}{\alpha}$, we have :

\begin{equation}\label{22l}
\Omega_{2}=-\frac{\upsilon_{\varphi}}{2}\alpha^{-1}.
\end{equation}

\begin{figure}[H]
\centering
\includegraphics[width=3.5in]{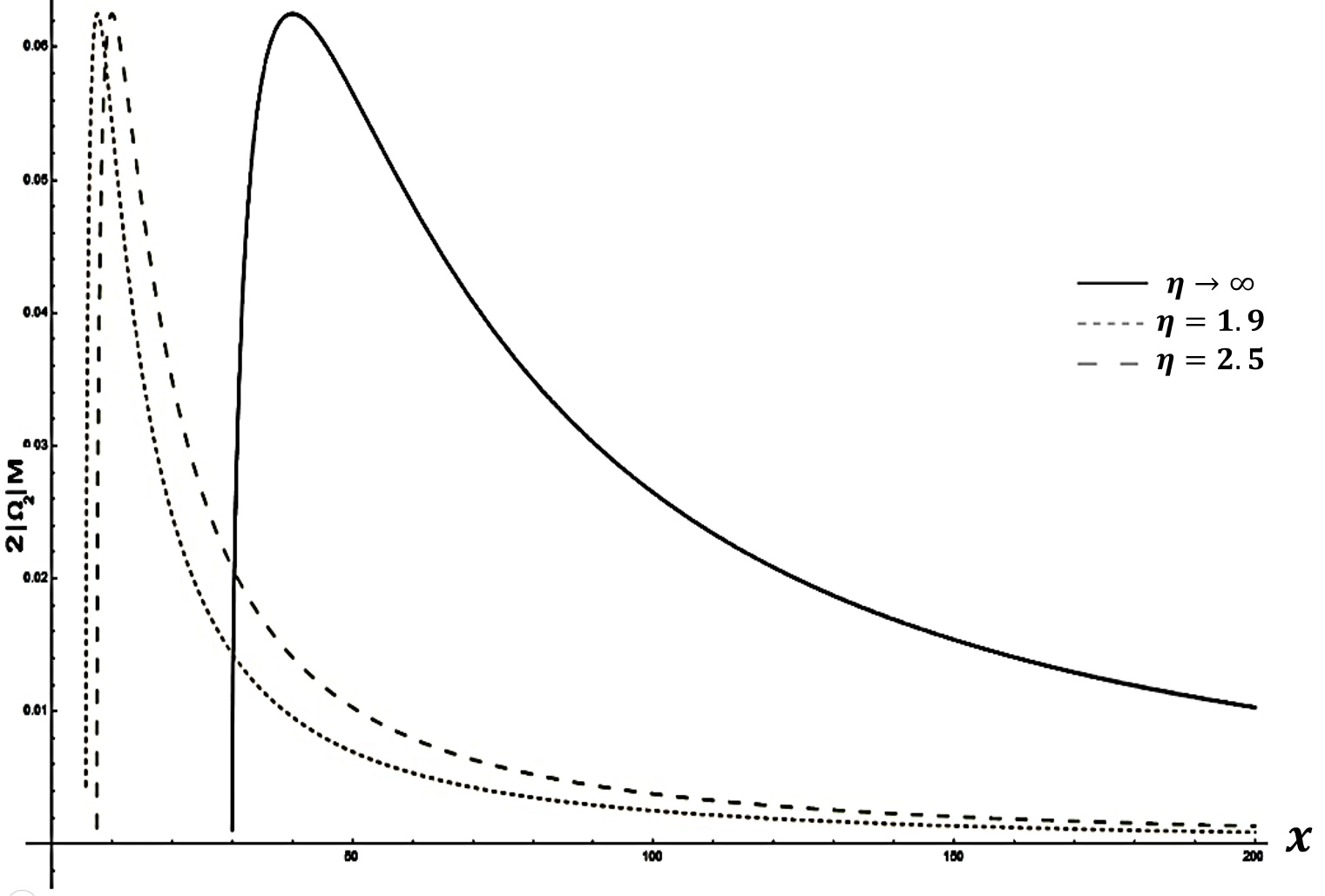}
\caption{\scriptsize  Neutrino spin oscillations frequency for  Schwarzschild metric vs. $ x $ for different values of $\eta$ in 3 dimension.\label{Fig1}}
\end{figure}

We have plotted $2|\Omega_{2}|M$ versus $x$ in Fig. \ref{Fig1}  in three dimensions for the Schwarzschild metric with different values of $\eta$.  By reducing the value of $\eta$ which corresponds to  increasing the noncommutative parameter, the peak of the curve occurs in shorter distances and it is observed that there is no big difference in the maximum  of the spin oscillation frequency with the increasing of the noncommutative effects. It is also seen that, as $\eta$ decreases, the spin oscillation tends to zero faster. This  is in agreement with the result we got before: the non-commutativity of space has an opposite role to the curvature of space.

\begin{figure}[H]
\centering
\includegraphics[width=4.5in]{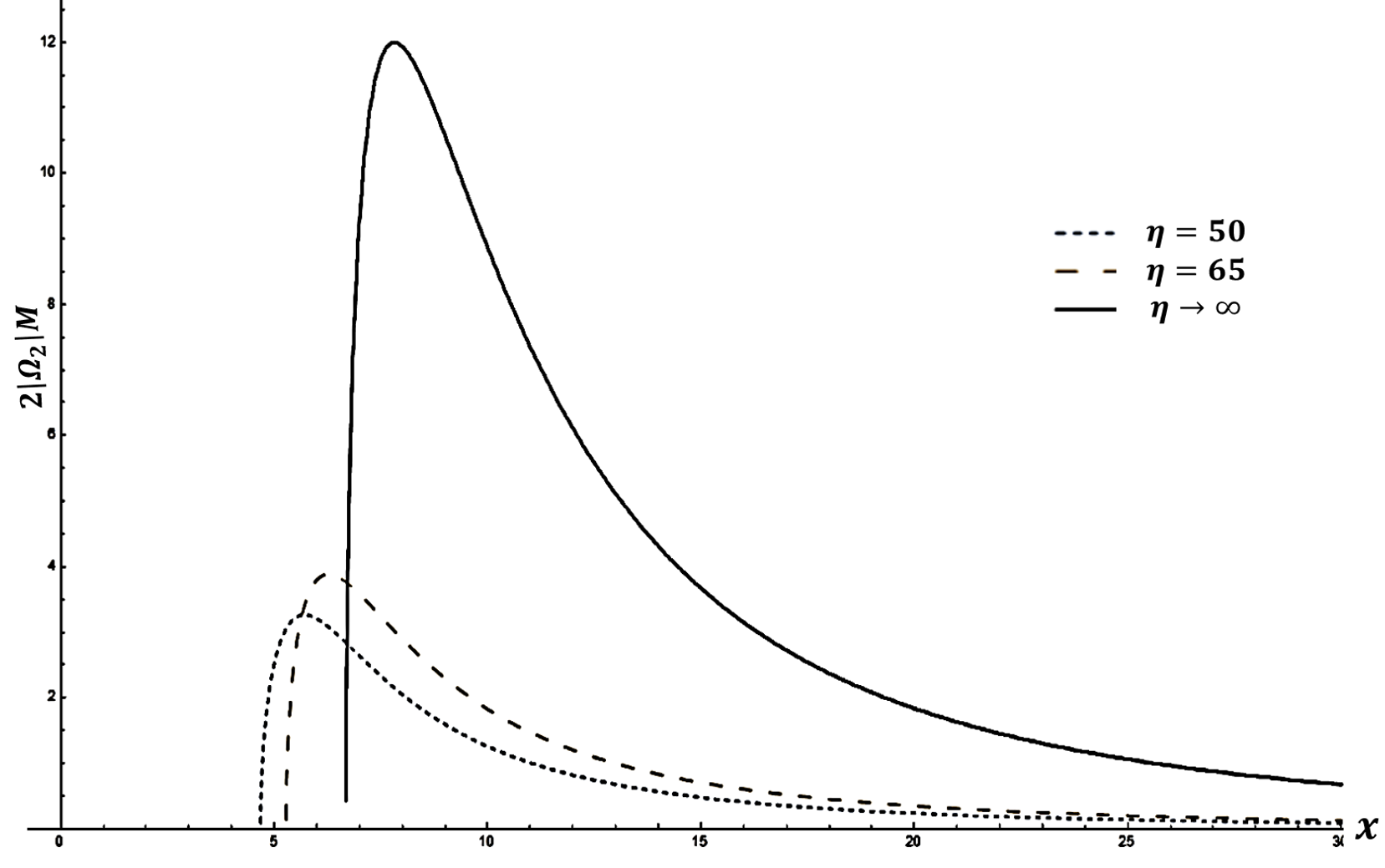}
\caption{\scriptsize  Neutrino spin oscillations frequency in Schwarzschild metric  vs. $ x $ for different values $\eta$ in 5 dimensions.\label{Fig2}}
\end{figure}

 In Fig. \ref{Fig2}, we have shown the oscillation  frequency of  neutrino spin for the Schwarzschild metric in five dimensions  with different $\eta$ and we observe that,  increasing the  non-commutativity of space  in higher dimensions reduces the peak of the spin oscillation frequency. It can also be checked that, by increasing the non-commutativity of space, the spin oscillations tend to zero, faster than commutative space which confirms the result we got earlier for the case of three dimensions.

\begin{figure}[H]
\centering
\includegraphics[width=4in]{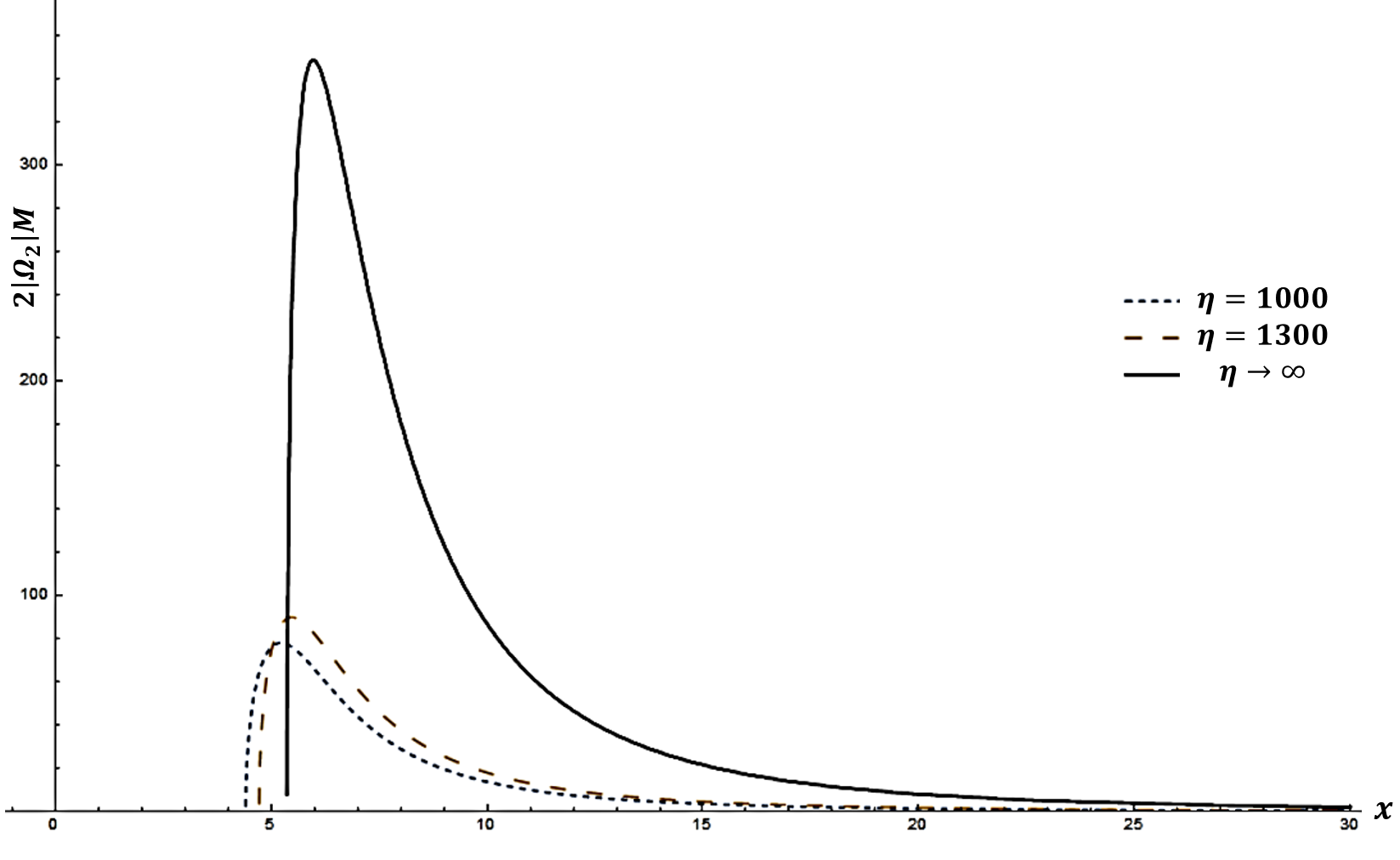}
\caption{\scriptsize  Neutrino spin oscillations frequency in  Schwarzschild metric vs. $ x $ for different values of $\eta$ in 7 dimensions.\label{Fig3}}
\end{figure}

In Fig. \ref{Fig3}, it can be seen that the frequency peak of the neutrino spin oscillation is decreased further with increasing the noncommutative effects and by comparing Figs. \ref{Fig2} and \ref{Fig3}, we observe that the reduction of the peak related to the increase in noncommutative effects is greater in seven dimensions than in five dimensions. This mean that by increasing the dimension of space, the effects of non-commutativity of space will become more obvious. In addition,  in seven dimension compared to five dimension, the peaks of the curves occur at  shorter distances. Furthermore, by increasing the non-commutativity of space, the amplitude of oscillations tends to zero more rapidly in agreement with the situation in three and five dimensions.

\begin{figure}[H]
\centering
\includegraphics[width=4in]{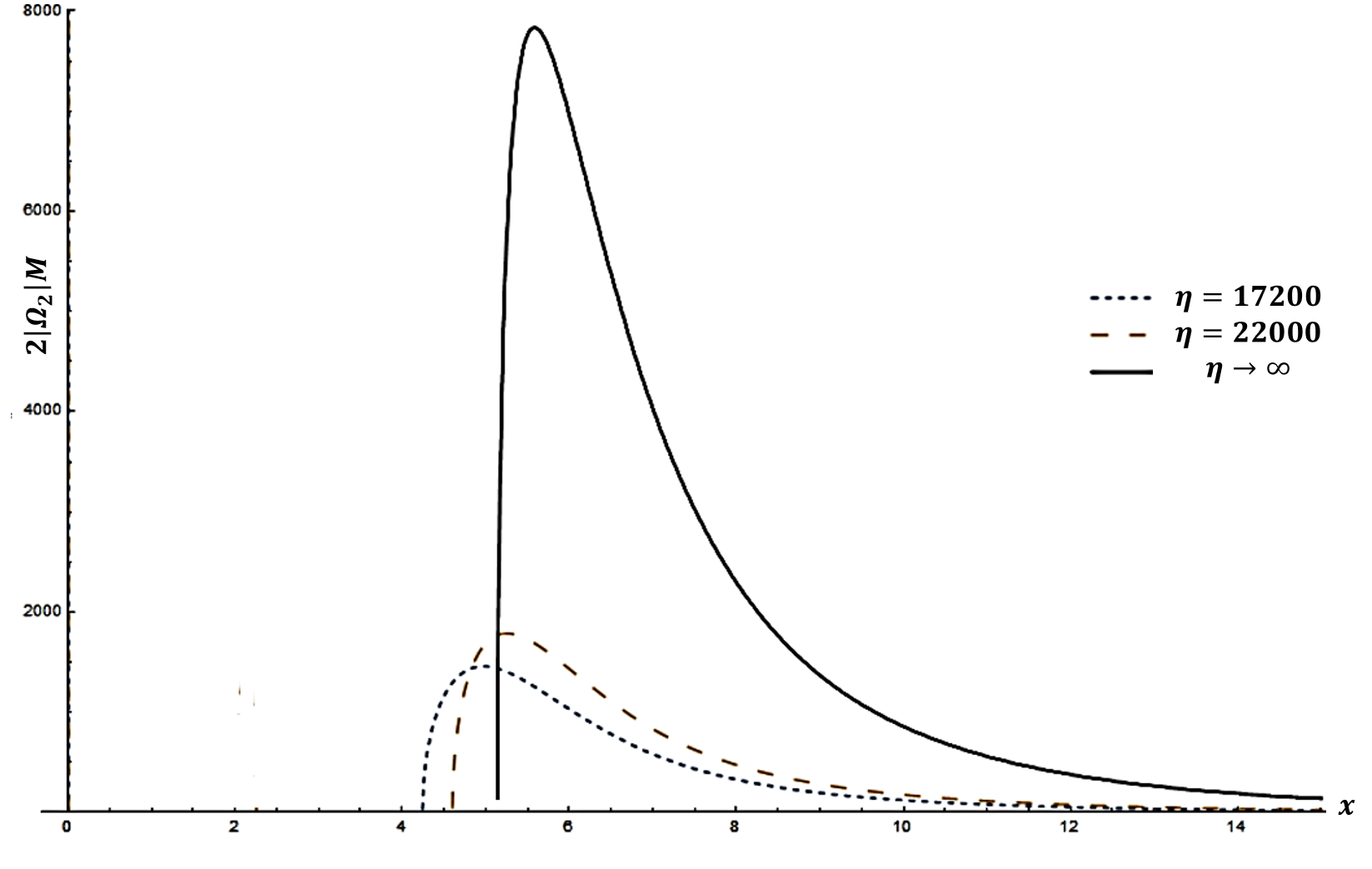}
\caption{\scriptsize Neutrino spin oscillations frequency  in Schwarzschild(Sh) metric  in  9 dimensions.\label{Fig15}}
\end{figure}

\begin{figure}[H]
\centering
\includegraphics[width=4in]{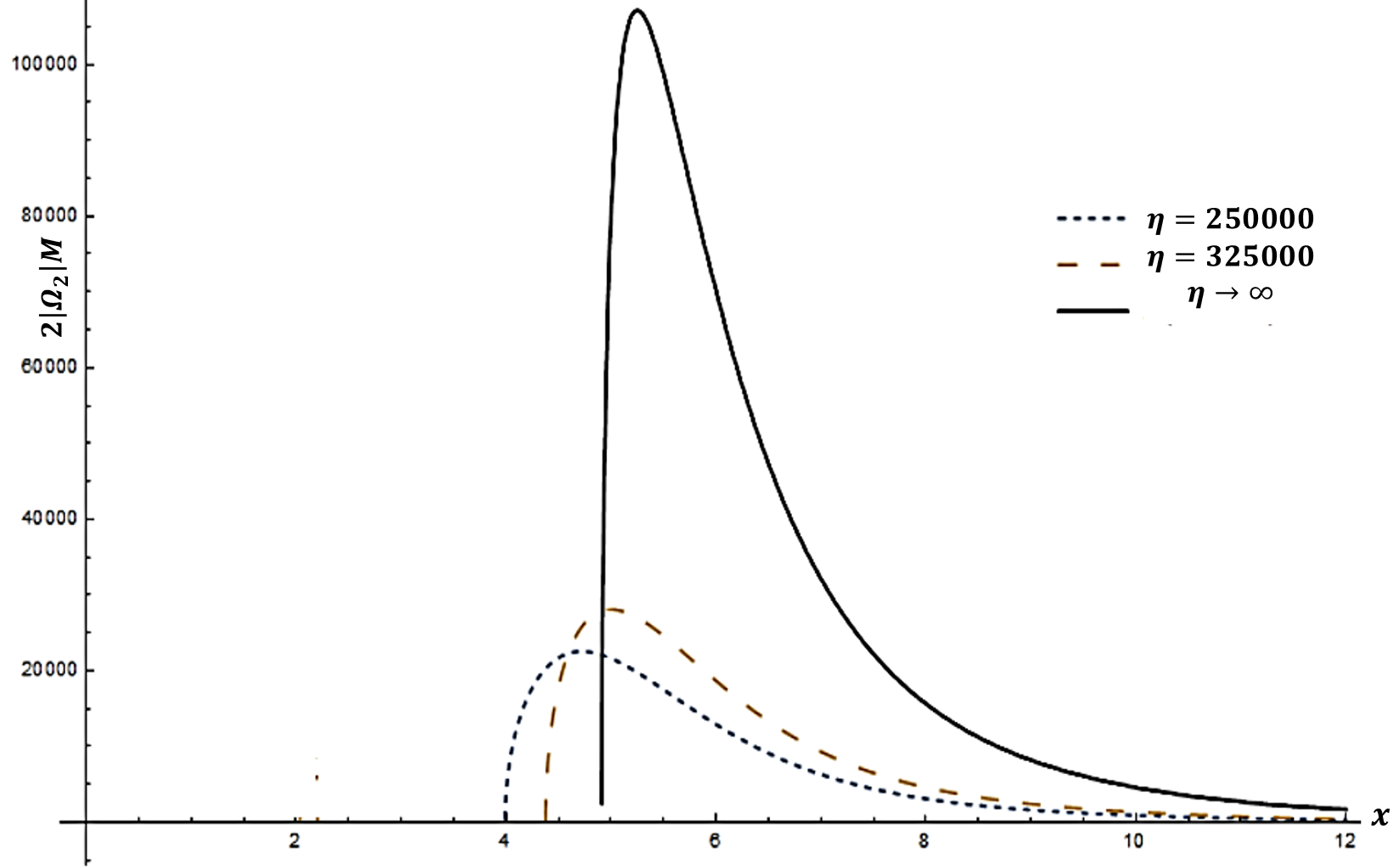}
\caption{\scriptsize  Neutrino spin oscillations frequency in Schwarzschild(Sh) metric  in 11 dimensions.\label{Fig16}}
\end{figure}
In Figs.  \ref{Fig15} and \ref{Fig16}, frequencies of neutrino spin oscillations vs. $x$  in nine and eleven dimensions are plotted  respectively. Similar to other dimensions, we observe that the reduction of $\eta$, or equivalently, the increase in the noncommutative parameter, leads to a decrease in the maximum value of the oscillation frequency. It is also observe that, by increasing the noncommutative effects (decreasing $\eta$), the spin oscillations tend to zero faster. So we conclude that, in each dimension, the non-commutativity of space plays an opposite character to the curvature of space.

\begin{figure}[H]
\centering
\includegraphics[width=4in]{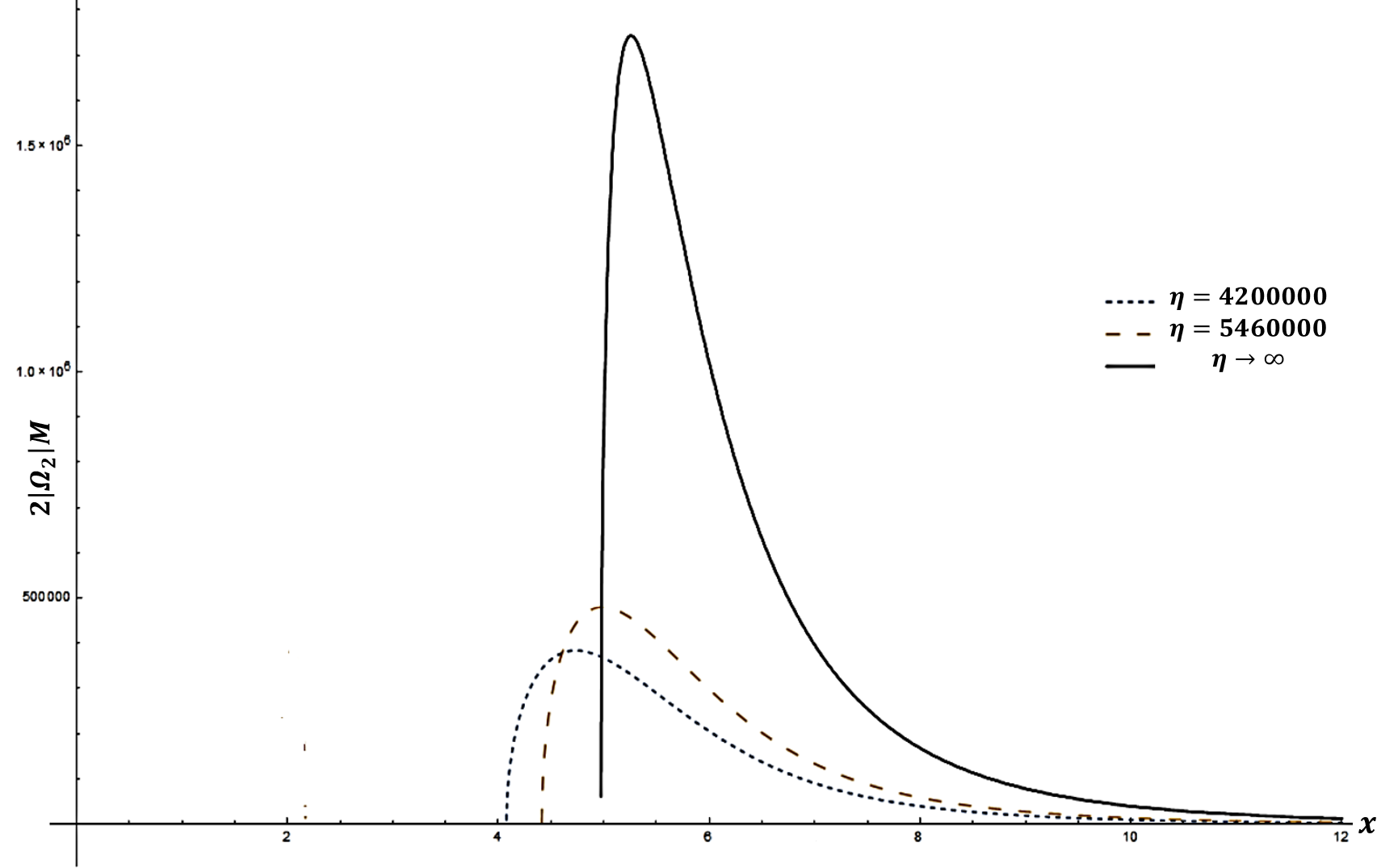}
\caption{\scriptsize  Neutrino spin oscillations frequency in Schwarzschild(Sh) metric  in 13 dimensions. \label{Fig30}}
\end{figure}

\begin{figure}[H]
\centering
\includegraphics[width=4in]{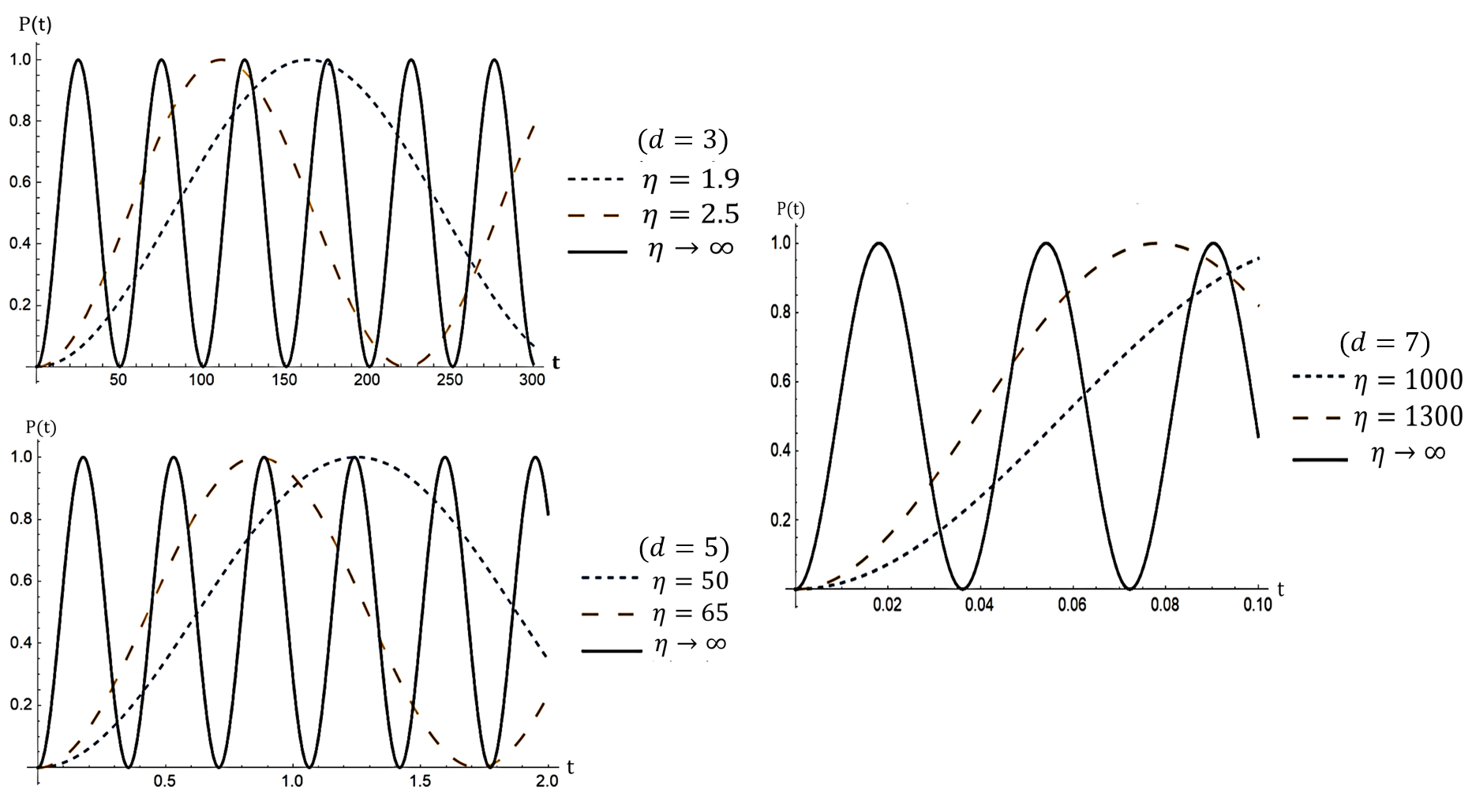}
\caption{\scriptsize  Probability of neutrino spin oscillations for Schwarzschild metric in 3-7 dimensions. \label{Fig33}}
\end{figure}

Now, with a simple check of the three diagrams in Figure \ref{Fig33}, we observe that:\\
\begin{equation}\label{290}
T^{d=3}_{\eta\rightarrow\infty}<T^{d=3}_{\eta=2.5}<T^{d=3}_{\eta=1.9},
\end{equation}
$$T^{d=5}_{\eta\rightarrow\infty}<T^{d=5}_{\eta=65}<T^{d=5}_{\eta=50}$$,
$$T^{d=7}_{\eta\rightarrow\infty}<T^{d=7}_{\eta=1300}<T^{d=7}_{\eta=1000}$$.

The relations (\ref{290}) show that in different dimensions as $\eta$ decreases, which is equivalent to an increase in the value of $\theta$, the period time of the neutrino spin oscillation probability increases. This could have significant observational implications, which will be discussed in the Phenomenological Applications section.

\section{Neutrino spin oscillation in non-commutative higher dimensional  Reissner-Nordström(RN) Metric }
The RN metric describes a black hole with charge Q and mass M. The  time component of its metric in the three dimensional non-commutative space is given by\cite{19,20} :
\begin{equation}\label{20}
A(r)=[1-\frac{4GM}{r\sqrt{\pi}}\gamma(\frac{3}{2},\frac{r^{2}}{4\theta})+\frac{Q^{2}}{\pi r^{2}}\gamma^{2}(\frac{1}{2},\frac{r^{2}}{4\theta})-\frac{Q^{2}}{\pi r \sqrt{2\theta}}\gamma(\frac{1}{2},\frac{r^{2}}{2\theta})]^{\frac{1}{2}},
\end{equation}
and in higher dimensions \cite{32} :
\begin{equation}\label{21}
A(r)=[1-\frac{
4GM}{r^{d-2}\pi^{\frac{d-2}{2}}}\gamma(\frac{d}{2},\frac{r^{2}}{4\theta})+\frac{(d-2)4Q^{2}G}{\pi^{d-3}r^{2d-4}}(\gamma^{2}(\frac{d}{2}-1,\frac{r^{2}}{4\theta})
\end{equation}
$$-\frac{2^{\frac{8-3d}{2}}r^{d-2}}{(d-2)\theta^{\frac{d-2}{2}}}\gamma(\frac{d}{2}-1,\frac{r^{2}}{2\theta}))]^{\frac{1}{2}},$$

by inserting  Eq. (\ref{20}) and Eq. (\ref{21}) in spin Oscillation equations introduced in the previous section , i.e. Eqs. (\ref{17l}),(\ref{18l}),(\ref{19l}),(\ref{20l}),(\ref{21l}) and by using Eq. (\ref{9l}), we get $\Omega_{2}$ in three and extra dimensions.

\begin{figure}[H]
\centering
\includegraphics[width=4in]{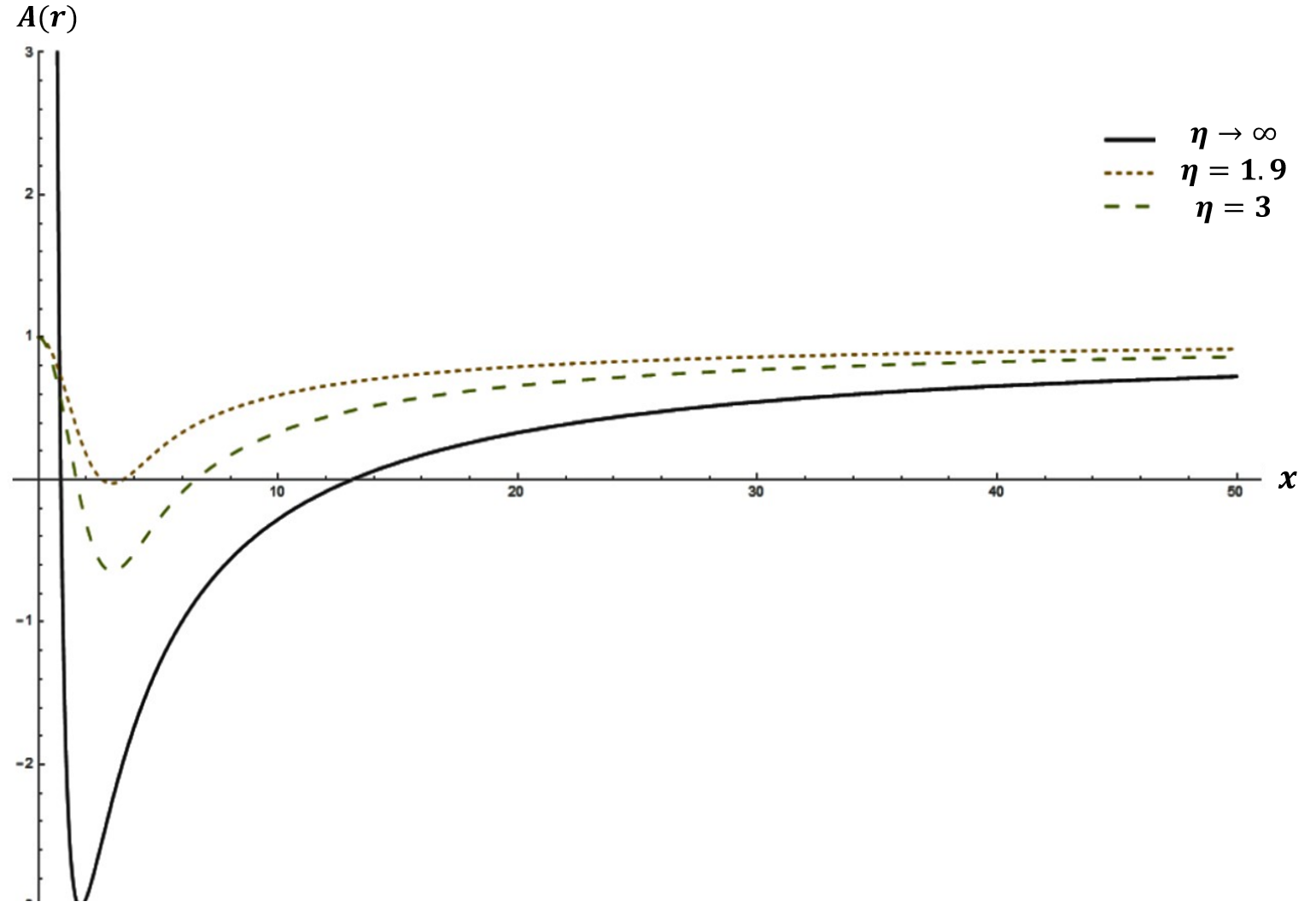}
\caption{\scriptsize The metric Rn vs. $ x $ for different values of $\eta$ in 3 dimensions.\label{Fig4}}
\end{figure}

 Fig. \ref{Fig4} shows the changes  of the RN metric in three dimensions versus x for three values of $\eta$.

$\eta\longrightarrow\infty$ represents the case  of commutative space. The intersections points with the x axis indicate the formation of a black hole and the existence of  event horizons. The metric approaches the metric of  the flat space in long distances, and we notice that this process happens faster in the NC space than  the commutative space.

In Figures (10)-(12), $2|\Omega_{2}|M$ in three, five and seven dimensional RN metric are represented as  functions of x for different values of $\eta$ respectively.
 The maximum values (peaks) reduce by decreasing $\eta$ and the peaks of the frequencies take place in smaller x as the value of NC parameter increases. As we get away from the horizon the values of frequencies go faster to zero by increasing  noncommutative effects (parameter), which means that, in long distances , the dominate contribution belongs to commutative space, but in short distance(small x) in higher dimensions, the contributions of noncommutative space become stronger.
\begin{figure}[H]
\centering
\includegraphics[width=3.5in]{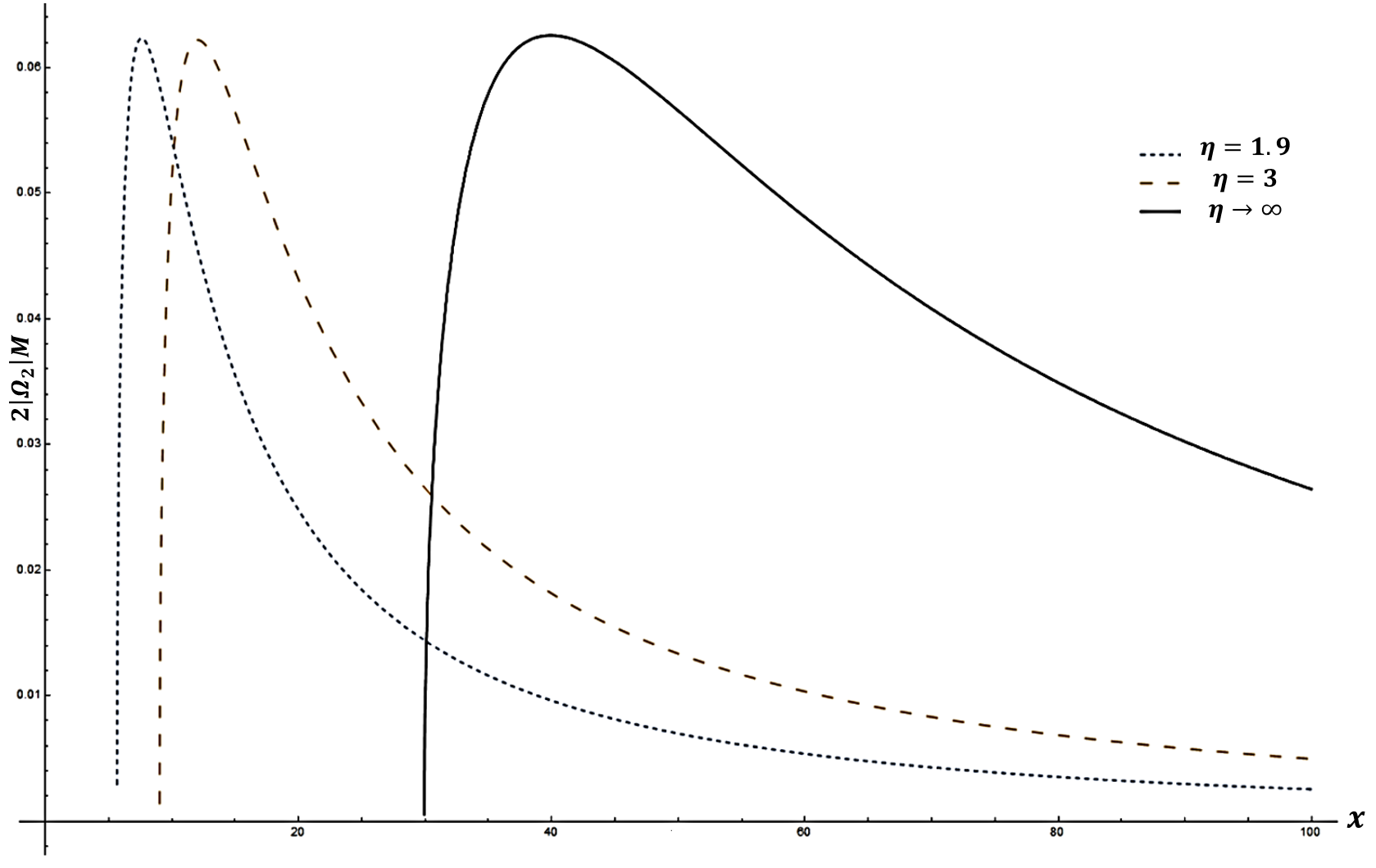}
\caption{\scriptsize $2|\Omega_{2}|M$ in RN metric vs. $ x $ for different values  of $\eta$ in 3 dimensions.\label{Fig32}}
\end{figure}

\begin{figure}[H]
\centering
\includegraphics[width=3.5in]{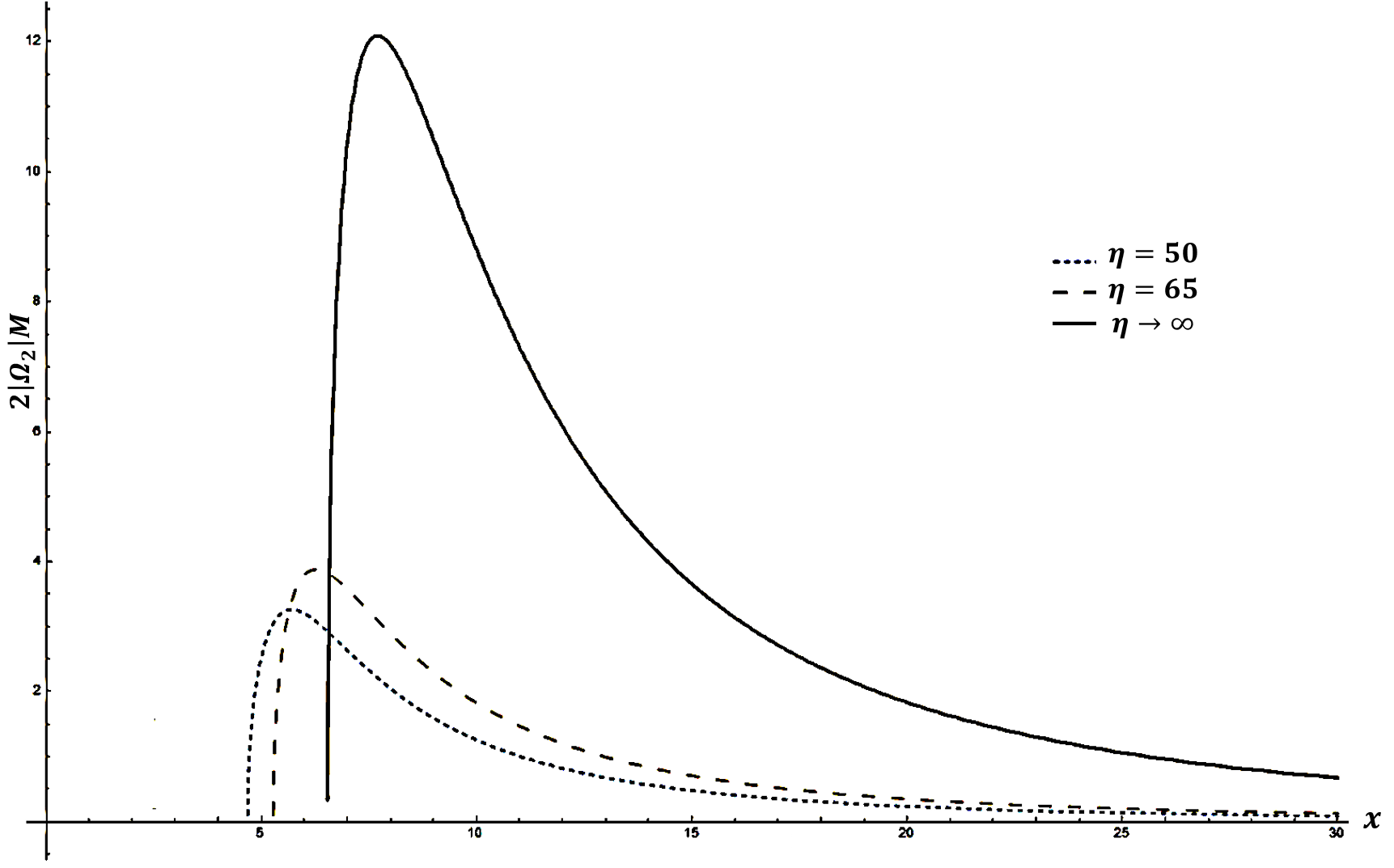}
\caption{\scriptsize $2|\Omega_{2}|M$ in RN metric vs. $ x $ for different values  of $\eta$ in 5 dimensions.\label{Fig5}}
\end{figure}

\begin{figure}[H]
\centering
\includegraphics[width=3.5in]{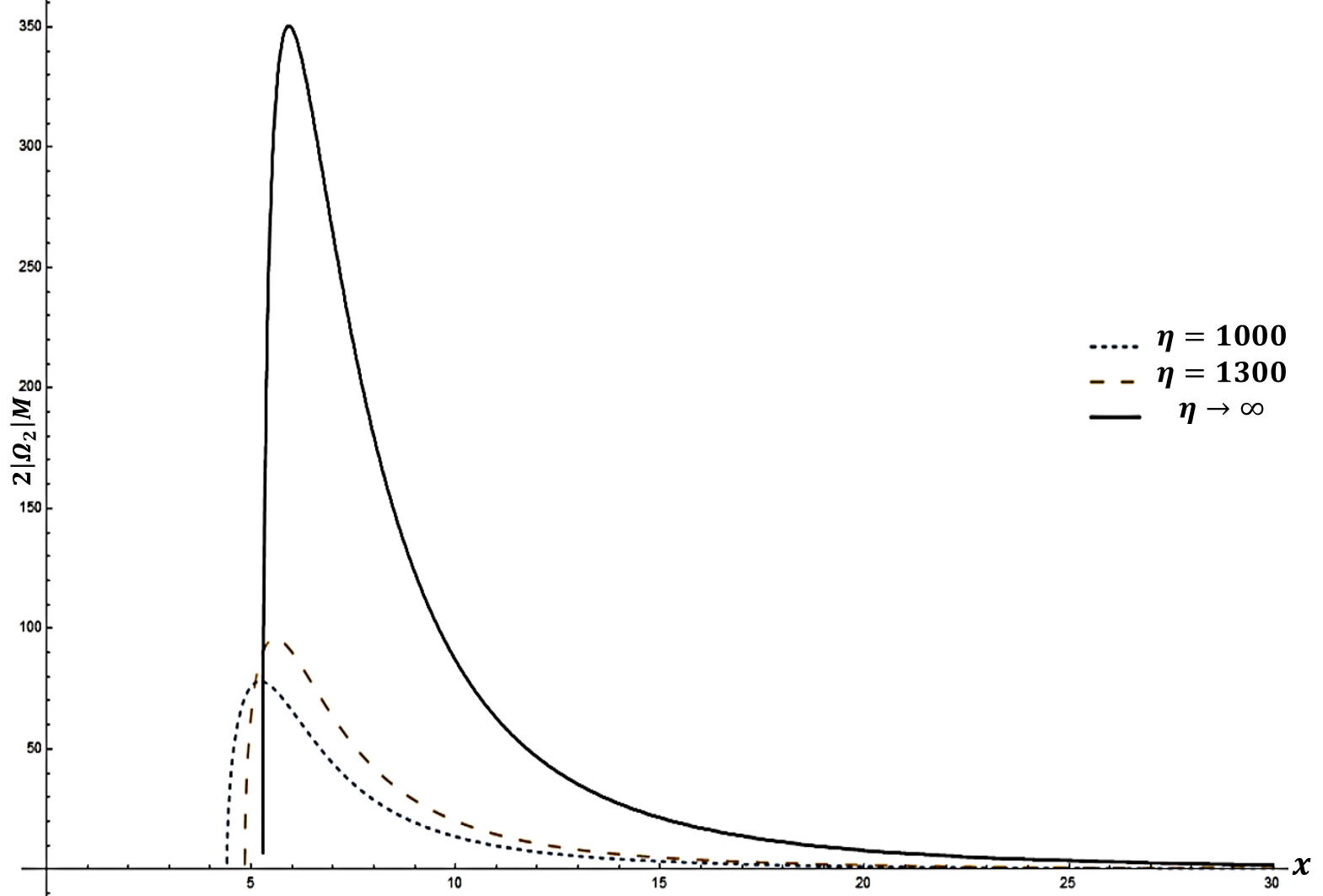}
\caption{\scriptsize $2|\Omega_{2}|M$ in Rn metric vs. $ x $ for different values of $\eta$ in 7 dimensions.\label{Fig6}}
\end{figure}

\begin{figure}[H]
\centering
\includegraphics[width=3.5in]{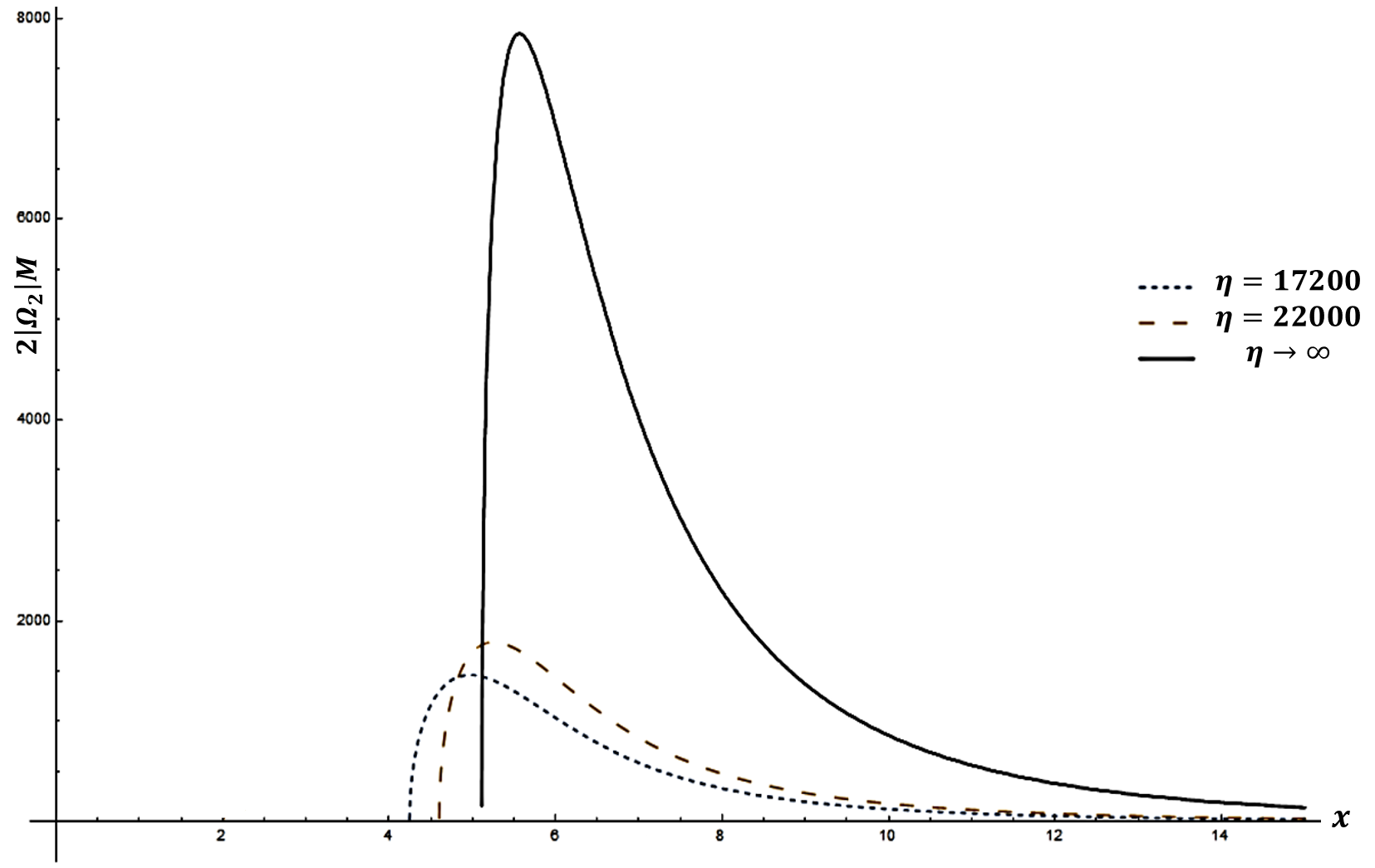}
\caption{\scriptsize  Neutrino spin oscillations frequency in RN  metric  for 9 dimensions($d$). \label{Fig13}}
\end{figure}

\begin{figure}[H]
\centering
\includegraphics[width=3.5in]{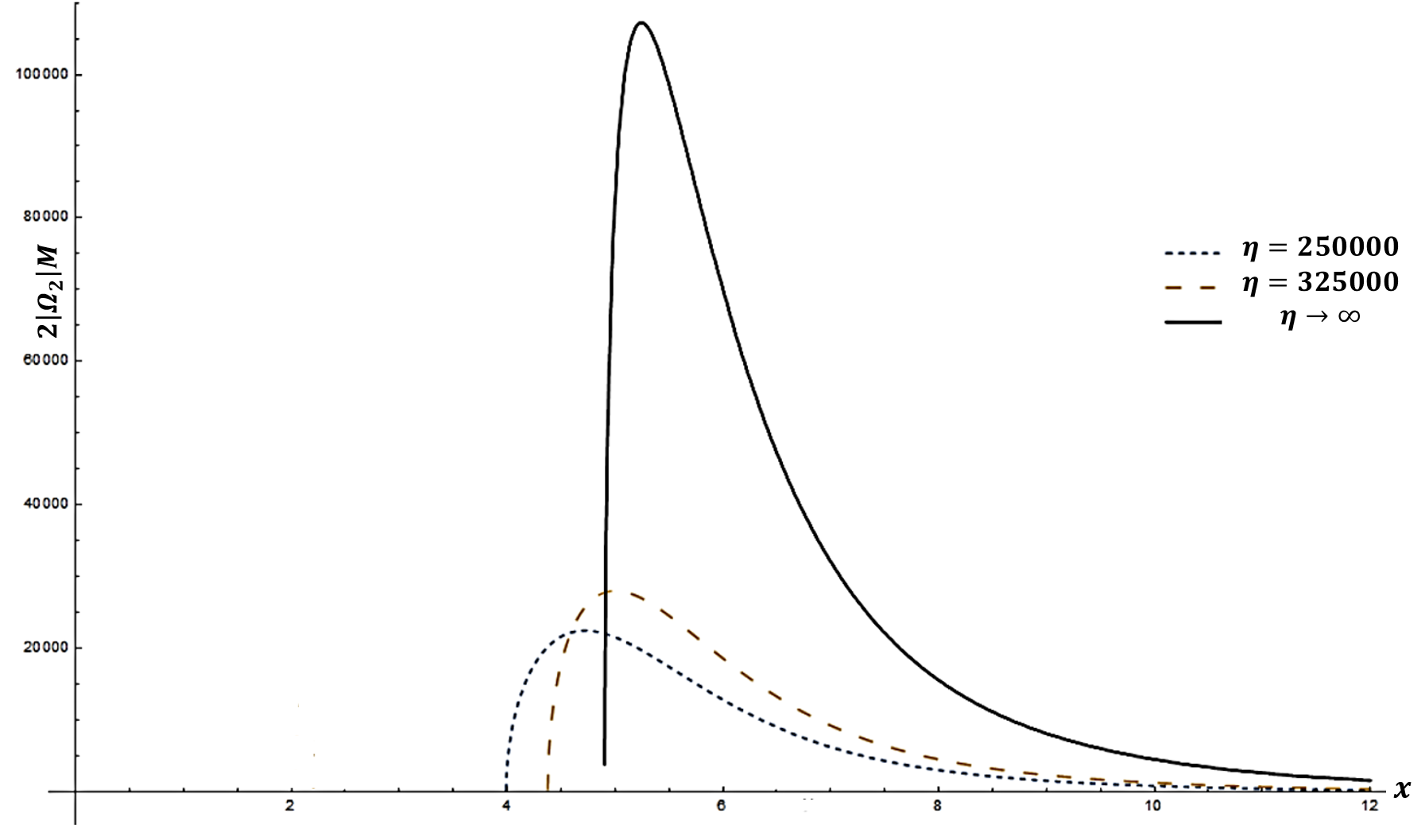}
\caption{\scriptsize  Neutrino spin oscillations frequency  in RN  metric  for 11 dimensions($d$). \label{Fig14}}
\end{figure}
We have also plotted $2|\Omega_{2}|M $ in nine, eleven and thirteen dimensions in Figures (13), (14) and (15) respectively. It can be seen that the peaks decrease as the noncommutative parameter  increases. By comparing the curves, it  is observed that  the difference between  neutrino spin oscillations frequencies in  non-commutative spaces compared to commutative cases, become smaller.

\begin{figure}[H]
\centering
\includegraphics[width=3.5in]{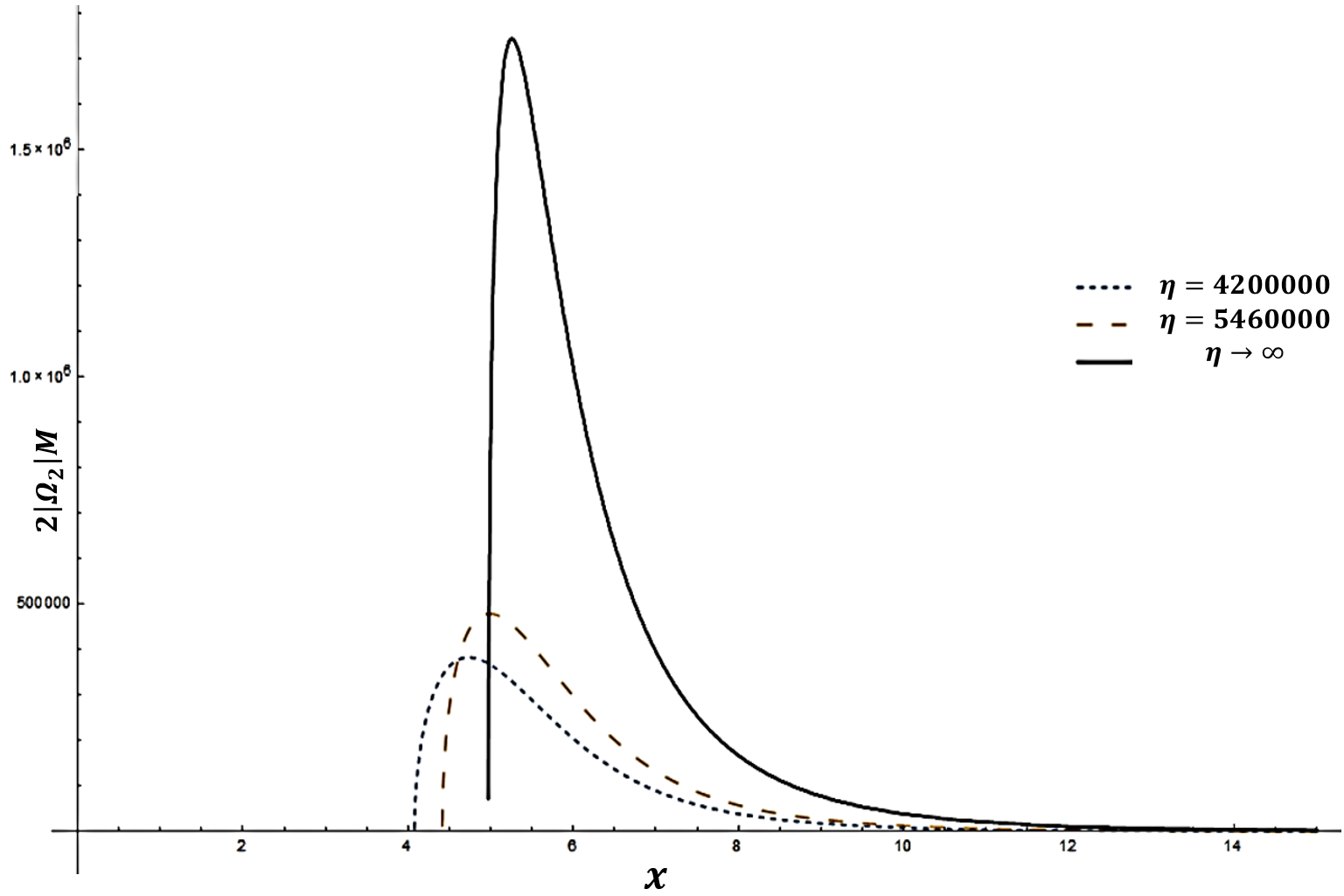}
\caption{\scriptsize  Neutrino spin oscillations frequency  in RN  metric  for 13 dimensions($d$). \label{Fig31}}
\end{figure}

\begin{figure}[H]
\centering
\includegraphics[width=5in]{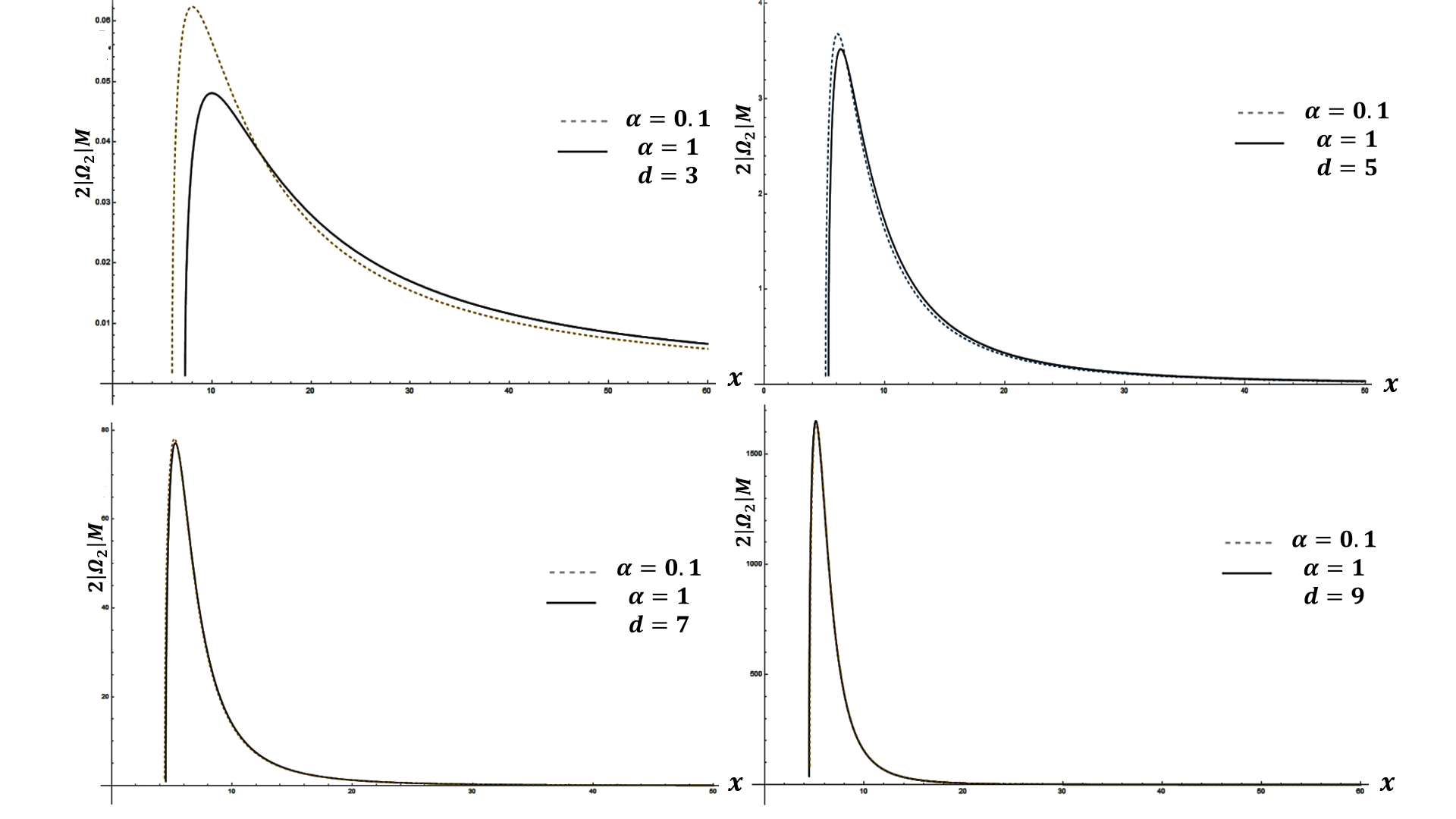}
\caption{\scriptsize The effects of the BH charge on neutrino spin oscillation frequency for RN metric vs. $ x $ for certain value of  $\eta$ in different dimensions.\label{Fig7}}
\end{figure}

Now, let us study the effects of black hole charge on the neutrino spin oscillations in noncommutative higher dimensional spaces.  To proceed further, we define new variables $\beta=\frac{2\sqrt{\pi}Q}{\sqrt{\theta}}$ and $\alpha=\frac{2\sqrt{\pi}Q}{M}$. Fig .\ref{Fig7}, shows that in all dimensions, black hole charge has a negative effects on the spin oscillations i.e., by increasing the charge of the black hole in all dimensions, the amplitude of spin oscillations decreases. It also shows that in noncommutative spaces, by increasing the dimension of space, the effects of black hole charge decreases which is the same as commutative spaces  \cite{33}.

\begin{figure}[H]
\centering
\includegraphics[width=5in]{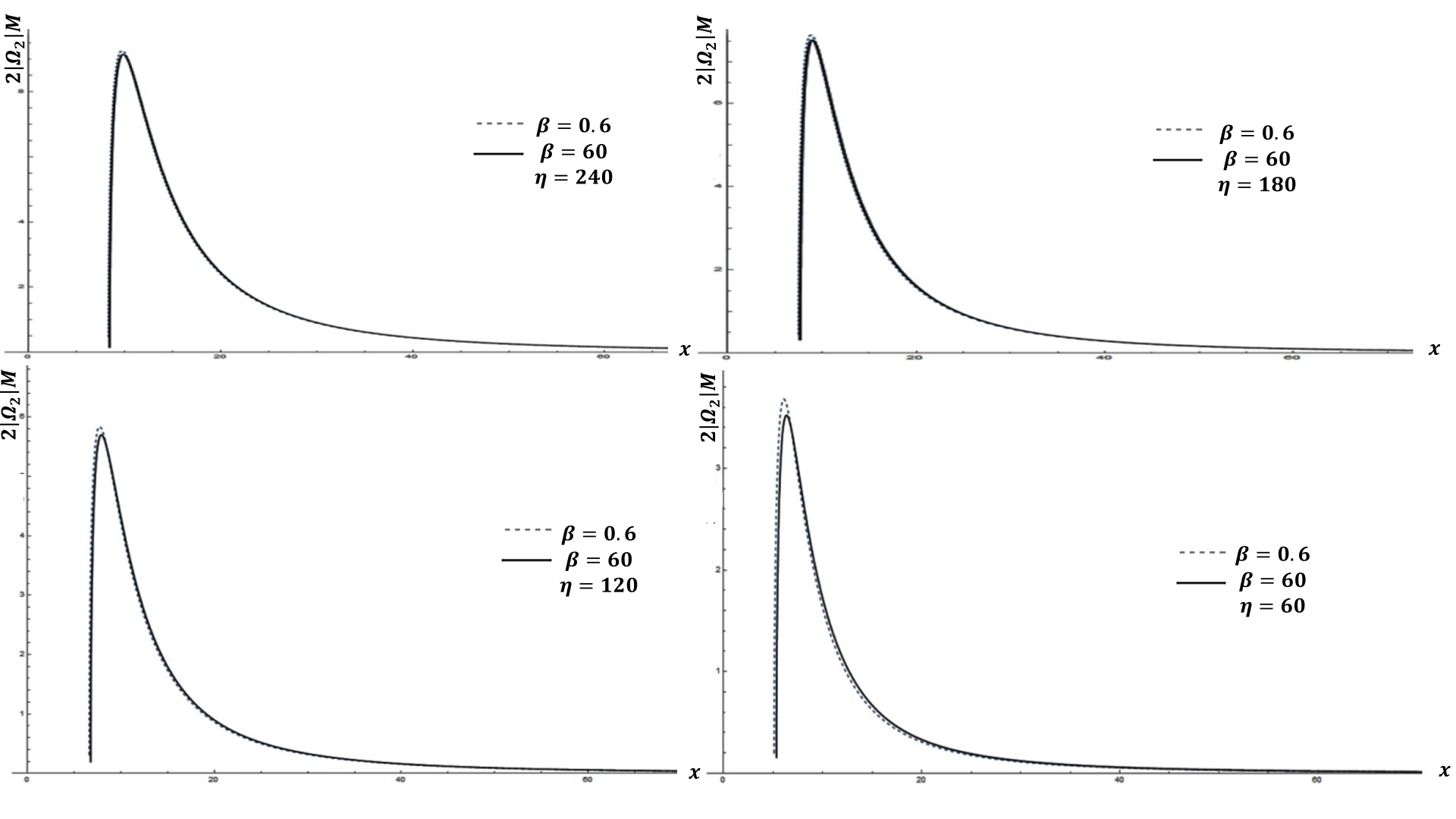}
\caption{\scriptsize The effects of the BH charge on neutrino spin oscillation frequency in RN metric  vs. $ x $ for different values of $\eta$ in 5 dimensions.\label{Fig8}}
\end{figure}

Fig. \ref{Fig8} indicates that, in an arbitrary dimension in the noncommutative case, increasing the non-commutativity of space makes the black hole charge more pronounced.

\begin{figure}[H]
\centering
\includegraphics[width=5.5in]{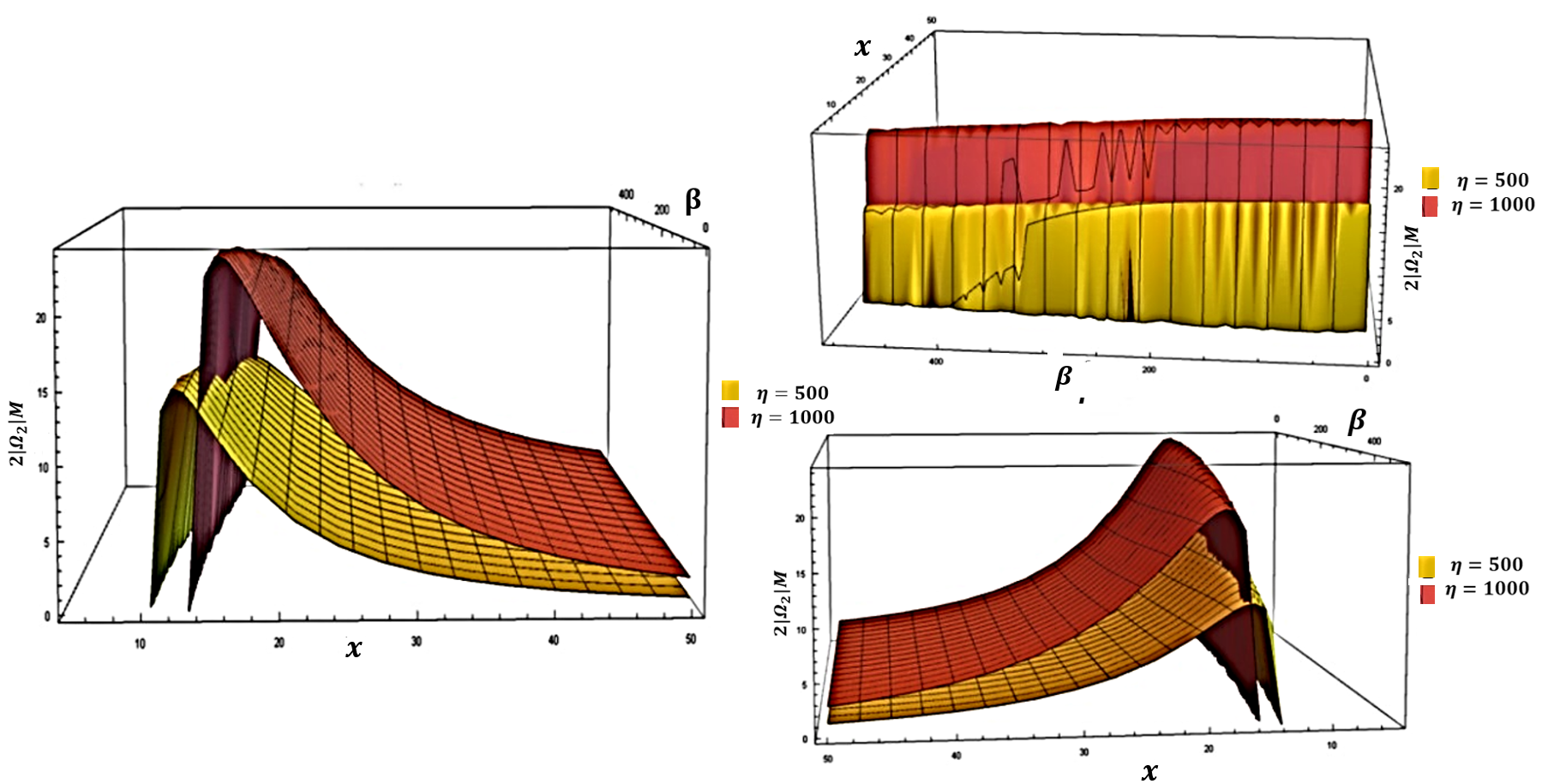}
\caption{\scriptsize The effects of the black hole charge on neutrino spin oscillation frequency for RN metric  vs. $ x $  with different values of $\eta$ and $\beta$ in 5 dimensions.\label{Fig10}}
\end{figure}

\begin{figure}[H]
\centering
\includegraphics[width=5in]{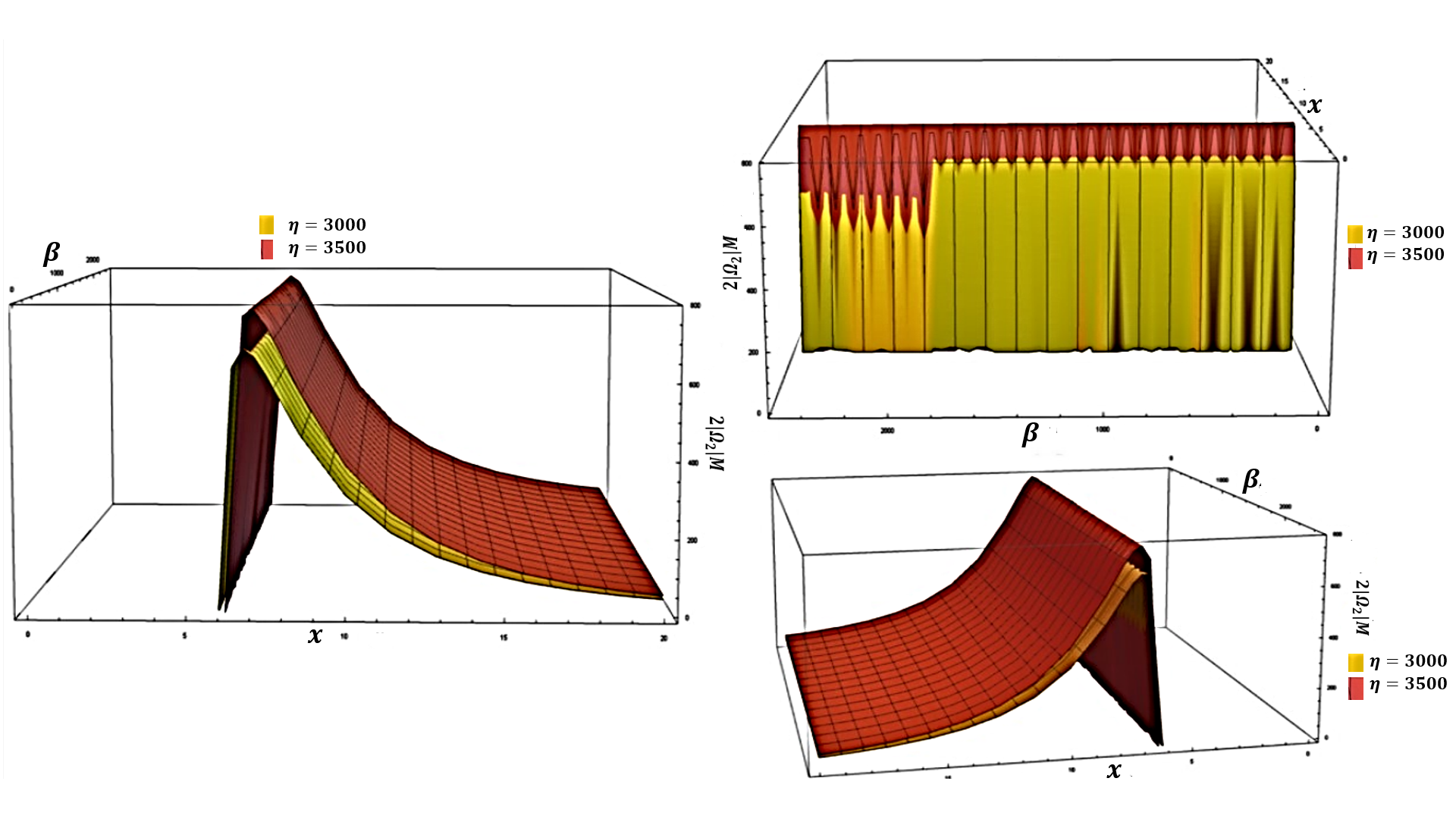}
\caption{\scriptsize The effects of the black hole charge on neutrino spin oscillation frequency for RN metric  vs. $ x $ with different values of $\eta$ and $\beta$ in 7 dimensions.\label{Fig11}}
    \end{figure}

In Fig. \ref{Fig10}, $ 2|\Omega_{2}|M $ is shown  for two values of $\eta=500 $ and  $\eta=1000 $ in  5 dimensions and in Fig. \ref{Fig11} for $\eta=3000 $ and $\eta=3500 $ in 7 dimension. By increasing the noncommutative effects which  occur in smaller $\eta$, and by  increasing the charge of the black hole, we can see a decrease in the peak of the spin oscillations curve in both figures \ref{Fig10} and  \ref{Fig11}, although we see a smaller difference in 7 dimensions (Fig. \ref{Fig11}).
Similar to the Schwarzschild black hole, we can examine the effect of space non-commutativity on the probability of neutrino spin oscillation in different dimensions for the RN black hole by plotting the relevant graphs. However, we will only mention the results, which indicate that, similar to the Schwarzschild black hole, with an increase in space non-commutativity, the oscillation period increases.

\section{Upper bounds on non-commutativity parameter }

Finally we can impose some upper bounds on non-commutativity parameter $\theta$. Using the fact that if the metric curve does not touch the x-axis  for some ranges of  $\eta$ values, so no black hole forms in this range, it is possible to find an upper bound on $\sqrt{\theta}$, which has an inverse relation with $\eta$. Upper bound on non-commutative parameter  have shown in Table(\ref{table12}) for higher dimensions in RN and Schwarzschild metrics respectively.\\\\

\begin{table}[H]
\centering
\includegraphics[width=4in]{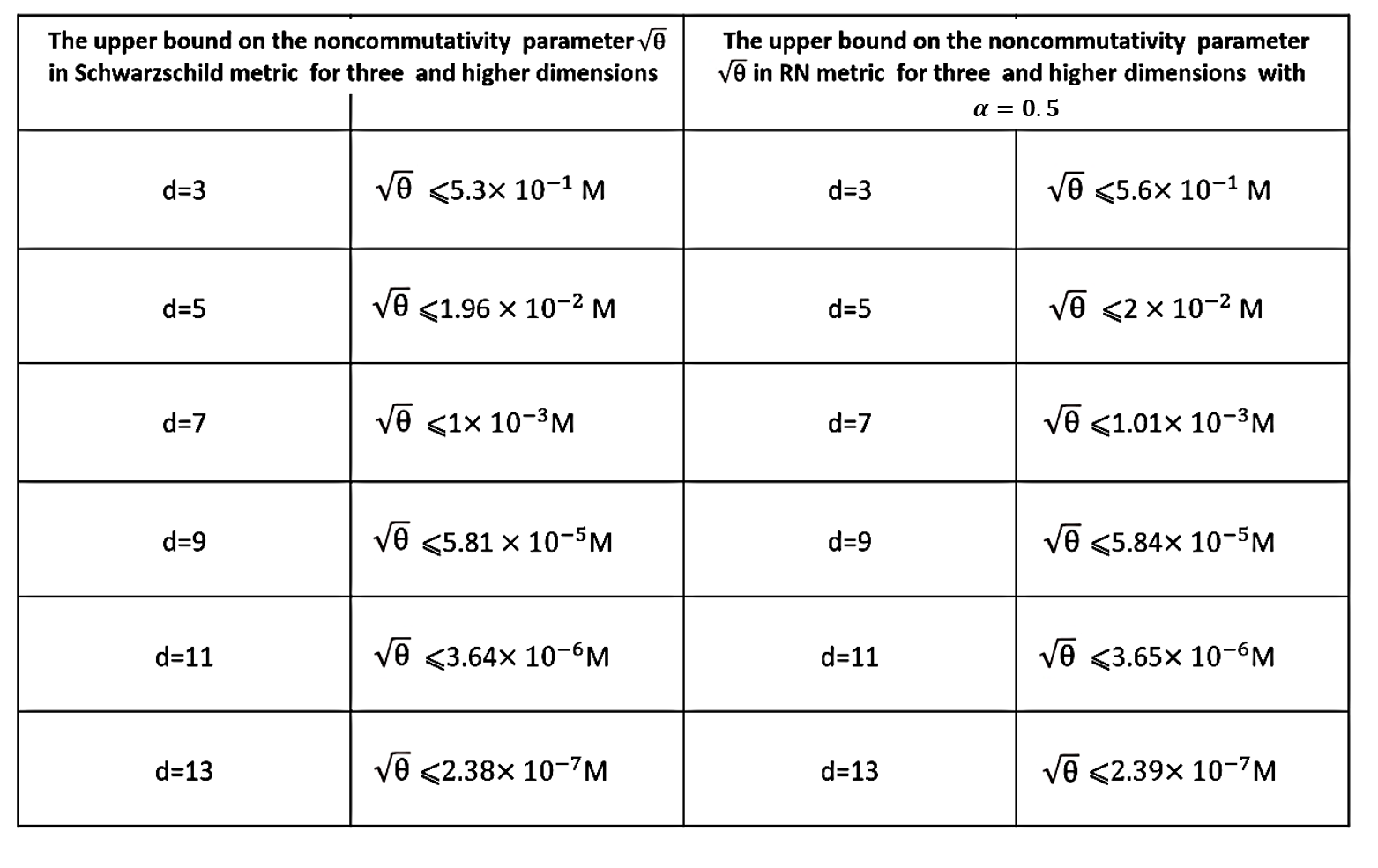}
\caption{\scriptsize Upper bound on $\sqrt{\theta}$ in RN and Schwarzschild metrics  for different dimensions($d$). \label{table12}}
\end{table}

Given the mass range of primordial black holes between $10^{-19}-10^{4}M_{\bigodot}$ \cite{40}, the most stringent upper bounds obtained from our calculations in Table\ref{table12} are summarized in Table(\ref{table13}).

\begin{table}[H]
\centering
\includegraphics[width=2in]{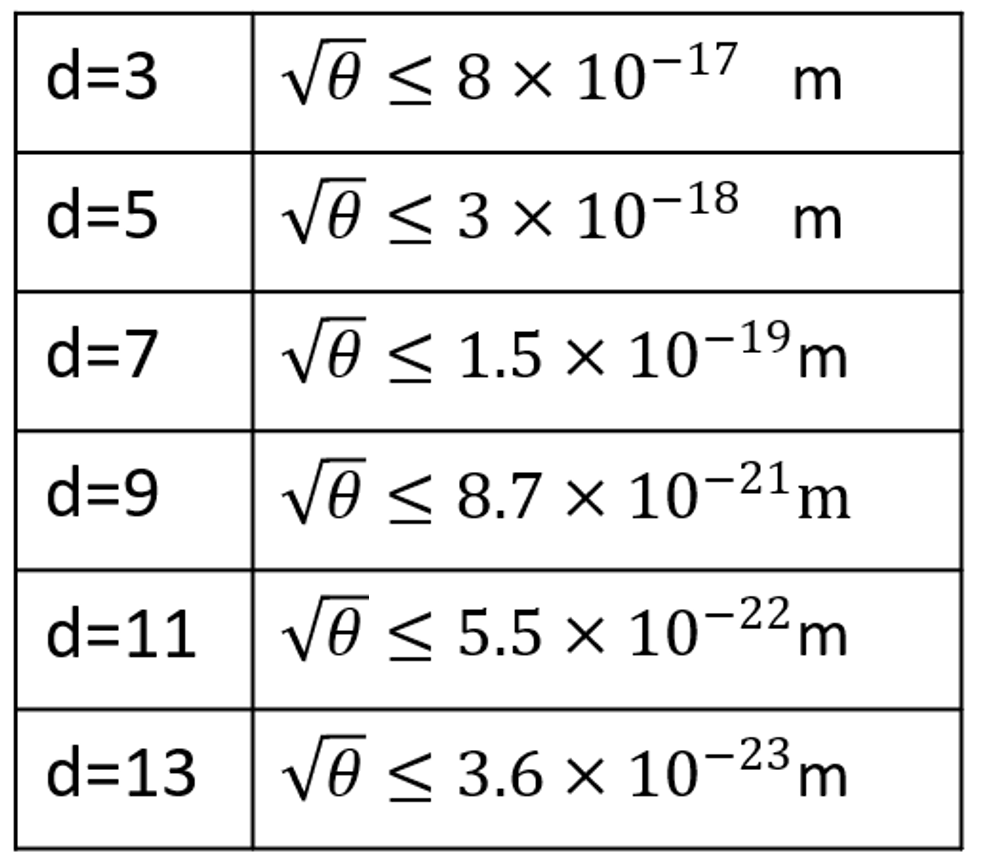}
\caption{\scriptsize Upper bound on $\sqrt{\theta}$ in RN and Schwarzschild metrics  for different dimensions($d$).\label{table13}}
\end{table}

Now, let us make a comparison between the upper bounds we have obtained for the non-commutativity parameter $\theta$ and existing theoretical or experimental constraints on this parameter. Some current bounds on non-commutative energy scale $\lambda^{-2}_{NC}\sim\mid\theta\mid$(Ref. \citen{41})are provided in Table(\ref{table14}).

\begin{table}[H]
\centering
\includegraphics[width=3in]{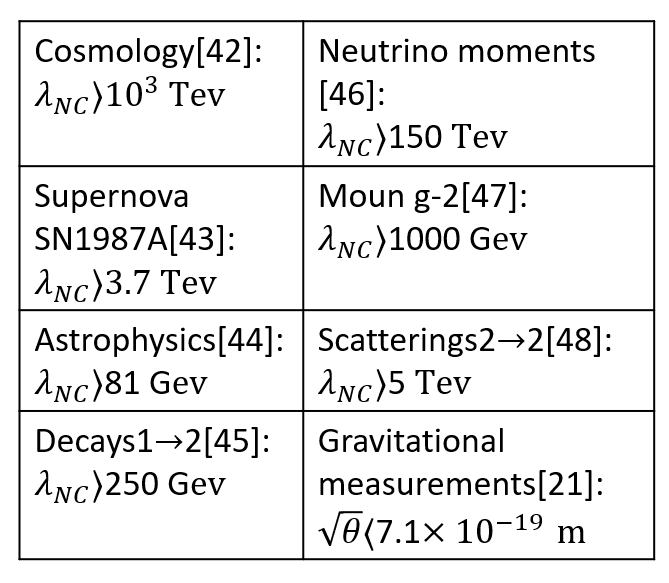}
\caption{\scriptsize Some current bounds on $\sqrt{\theta}$ or $\lambda$, the numbers in the brackets show the number of related references. \label{table14}}
\end{table}
The most stringent upper limit for $\theta$ obtained from the above table corresponds to $\lambda_{NC}>10^{3}$ TeV which yields the following value $\sqrt{\theta}\leq10^{-11} m$. It is evident that the limits obtained for $\theta$ in Table(\ref{table13}) are much more stringent than this limit.

It is observed that, stricter bound on $\theta$ are obtained in higher dimensions. It is worth mentioning that a lower bound for NC parameter $\theta$ has been obtained in some papers through the study of black holes thermodynamics in NC spaces \cite{18,19,20}.  It is also  interesting  to note that an upper bound on the dynamical noncommutative parameter has been obtained in Ref.\citen{49}.
\section{Phenomenological application}
Neutrinos in compact-object mediums such as core-collapse supernovae are exposed to different kinds of collective effects including bipolar collective oscillations, which can influence the dynamics and nucleosynthesis of those astrophysical environments. An example of such oscillations occurs in a system composed of equal amounts of neutrinos and antineutrinos. The neutrinos and antineutrino forms a bipolar vector in flavor isospin space. Now let us consider a simple neutrino bipolar system which helps us to understand many of the qualitative features of the collective neutrino oscillations in super-novae. The system composed of a homogeneous and isotropic gas that initially consists of mono-energetic neutrinos and antineutrinos and is described by the flavor pendulum. We introduce $\varepsilon$ as the fractional excess of neutrinos over antineutrinos. Now, let us examine such systems in the context of a neutral (Schwarzschild) black hole with the same initial value of $\varepsilon$ but with different values of $\eta$ in three dimension:
\begin{equation}\label{210}
(\varepsilon^{d=3}_{\eta\rightarrow\infty})_{t=0}=(\varepsilon^{d=3}_{\eta=2.5})_{t=0}=(\varepsilon^{d=3}_{\eta=1.9})_{t=0}.
\end{equation}

Considering the relations (\ref{210}), which indicate that with the increase in non-commutativity of space, the period time of neutrino oscillation probability increases, so at a later time $t >0$, we will have:

\begin{equation}\label{211}
(\varepsilon^{d=3}_{\eta\rightarrow\infty})_{t}<(\varepsilon^{d=3}_{\eta=2.5})_{t}<(\varepsilon^{d=3}_{\eta=1.9})_{t}.
\end{equation}

The same result is obtained for the RN metric and for higher dimensions. This implies that the spin oscillations frequencies of the flavor pendulum as a function of the neutrino number density for both metrics and in any dimension, decreases with the increase in spatial non-commutativity. Therefore, a detailed study of neutrino fluxes could serve as a tool for exploring space non-commutativity and extra dimensions in the future.

\section{Discussion}
We have analyzed the dependence of the oscillation frequency on the orbital radius for both Schwarzschild and RN metrics in gravitational fields in noncommutative higher dimensional spaces. We found out that in both metrics the maximum of the frequencies decreases by increasing the non-commutativity of space and the peaks occur in smaller distances compared to the three dimensional case. It is also observed that for long distances the main contribution to the neutrino spin transition probability belongs to three dimension. We have also studied the effects of the black hole charge on the frequency of spin oscillations in noncommutative higher dimensions. It is shown that in all dimensions, by increasing the noncommutative effects the role of black hole charge on spin oscillations increases. More explanations are provided below:

a). In every arbitrary dimension, the peak of the spin oscillations frequency decreases with increasing non-commutativity of space. This means that by increasing the non-commutativity of space in any arbitrary dimension, the oscillation frequency generally decreases, leading to an increase in the oscillation period. This fact can also be investigated by plotting the oscillation probability as a function of time. This is done in Figure \ref{Fig33}, which clearly shows that with increasing space non-commutativity, the oscillation period increases. This is a very important result, as it implies that with an increase in the non-commutativity of space in any chosen dimension, the oscillation period of neutrino spin increases. Consequently, for example, if we consider neutrinos of the Majorana type, an increase in the non-commutativity of space reduces the conversion rate of neutrinos to antineutrinos. This could impact the collective behavior of neutrinos in environments such as super-novae.

b). In each dimension, as the non-commutativity of space increases, the peak shifts towards smaller r. Therefore, in each dimension, with an increase in the non-commutativity of space, gravity becomes weaker and can manifest its strength over shorter distances near the black hole. Therefore, if non-commutativity of space exists, its effects can be measured at distances closer to the black hole rather than at farther distances.

c). In each dimension, as the non-commutativity of space increases, the oscillations tend to zero faster. This is also consistent with the points mentioned above and demonstrates that with an increase in non-commutativity in each dimension, gravity becomes weaker and, as a result, approaches zero more quickly as the distance from the black hole increases.
As an important result, in every dimension, the non-commutativity of space plays a role contrary to the curvature of space and reduces gravity.\\
 The detection of gravitational waves is significant for several reasons, as it has opened up a new window to the universe and advanced our understanding of fundamental physics. Similarly, the detection of neutrinos and the study of neutrino spin oscillations can serve as highly useful tools in investigating the structure of space-time, addressing questions such as whether space can be non-commutative and the existence of extra dimensions. The ability of neutrinos to pass through matter far exceeds that of photons, so in the future, neutrino telescopes may potentially rival electromagnetic telescopes like the Event Horizon Telescope in the study of black hole surroundings.

\section{Conclusion}
In our previous paper we examined the effects of extra dimensions on neutrino spin oscillations \cite{33}. In the present work, we extend our calculations and studies to the noncommutative spaces, so in this paper, three important topics in physics are simultaneously involved: the structure of space-time and non-commutative spaces, neutrino oscillations, and extra dimensions. We have studied and analyzed the effects of space non-commutativity on the oscillation frequency and spin oscillation probability of neutrinos in various dimensions for Schwarzschild and Reissner-Nordström black holes. Understanding the underlying structure of
space-time is essential for formulating a theory of quantum gravity and potentially addressing some fundamental challenges including the cosmological constant problem.\\

{\textbf{\large Declaration of competing interest }}

The authors declare that they have no known competing financial interests or personal relationships that could have appeared to
influence the work reported in this paper.\\\\

{\textbf{\large Data availability }}

No data was used for the research described in the article.

\end{document}